\documentclass[USenglish,oneside,twocolumn]{article}

\usepackage[utf8]{inputenc}
\usepackage[big]{dgruyter_NEW}

\cclogo{\includegraphics{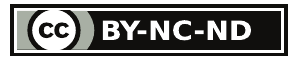}}

\usepackage{amsmath}
\usepackage{algorithmic}
\usepackage{url}
\usepackage{hyperref}
\usepackage{subcaption}
\usepackage{mathtools}
\mathtoolsset{showonlyrefs}

\usepackage{balance}

\newcommand{\sectionref}[1]{$\S$\ref{#1}}
\providecommand{\myparab}[1]{\smallskip\noindent\textbf{#1} }

\hypersetup{
	colorlinks	= true,
	citecolor	= black,
	filecolor	= black,
	linkcolor	= black,
	urlcolor	= black,
	pagebackref	= false,
}

\begin{document}

\author[1]{Nguyen Phong Hoang}
\author[2]{Arian Akhavan Niaki}
\author[3]{Phillipa Gill}
\author[4]{Michalis Polychronakis}

\affil[1]{Stony Brook University, E-mail: nghoang@cs.stonybrook.edu}
\affil[2]{University of Massachusetts - Amherst, E-mail: arian@cs.umass.edu}
\affil[3]{University of Massachusetts - Amherst, E-mail: phillipa@cs.umass.edu}
\affil[4]{Stony Brook University, E-mail: mikepo@cs.stonybrook.edu}

\title{\fontsize{20}{27}\selectfont Domain Name Encryption Is Not Enough:\\Privacy Leakage via IP-based Website Fingerprinting}

\runningtitle{Domain Name Encryption Is Not Enough: Privacy Leakage via IP-based Website Fingerprinting}

\begin{abstract}
{Although the security benefits of domain name encryption technologies such as
DNS over TLS (DoT), DNS over HTTPS (DoH), and Encrypted Client Hello (ECH) are
clear, their positive impact on user privacy is weakened by---the still
exposed---IP address information. However, content delivery networks,
DNS-based load balancing, co-hosting of different websites on the same server,
and IP address churn, all contribute towards making domain--IP mappings
unstable, and prevent straightforward IP-based browsing tracking.\\
In this paper, we show that this instability is not a roadblock (assuming a
universal DoT/DoH and ECH deployment), by introducing an IP-based website
fingerprinting technique that allows a network-level observer to identify
\emph{at scale} the website a user visits. Our technique exploits the complex
structure of most websites, which load resources from several domains besides
their primary one. Using the generated fingerprints of more than 200K websites
studied, we could successfully identify 84\% of them when observing solely
destination IP addresses. The accuracy rate increases to 92\% for popular
websites, and 95\% for popular \textit{and} sensitive websites. We also
evaluated the robustness of the generated fingerprints over time, and
demonstrate that they are still effective at successfully identifying about
70\% of the tested websites after two months. We conclude by discussing
strategies for website owners and hosting providers towards hindering IP-based
website fingerprinting and maximizing the privacy benefits offered by DoT/DoH
and ECH.}
\end{abstract}

\keywords{Domain Name Encryption, DoT, DoH, Encrypted Client Hello, Website
Fingerprinting}

\journalname{Proceedings on Privacy Enhancing Technologies}
\DOI{10.2478/popets-2021-0058}
\startpage{1}
\received{2021-02-28}
\revised{2021-06-15}
\accepted{2021-06-16}

\journalyear{}
\journalvolume{2021}
\journalissue{4}

\maketitle
\section{Introduction}

Due to the increase of Internet surveillance in recent
years~\cite{Fuchs2011InternetAS, Akbari2019PlatformSA}, users have become more
concerned about their online activities being monitored, leading to the
development of privacy-enhancing technologies. While various mechanisms can be
used depending on the desired level of privacy~\cite{Hoang2014:Anonymous},
encryption is often an indispensable component of most privacy-enhancing
technologies. This has led to increasing amounts of Internet traffic being
encrypted~\cite{EFF_ecrypt_theWeb}.

Having a dominant role on the Internet, the web ecosystem thus has witnessed a
drastic growth in HTTP traffic being transferred over
TLS~\cite{Felt2017MeasuringHA}. Although HTTPS significantly improves the
confidentiality of web traffic, it cannot fully protect user privacy on its
own when it comes to preventing a user's visited websites from being monitored
by a network-level observer. Specifically, under current web browsing
standards, the domain name information of a visited website can still be
observed through DNS queries/responses, as well as the Server Name Indication
(SNI) field of the TLS handshake packets. To address this problem, several
domain name encryption technologies have been proposed recently to prevent the
exposure of domain names, including DNS over TLS (DoT)~\cite{rfc7858}, DNS
over HTTPS (DoH)~\cite{rfc8484}, and Encrypted Client Hello
(ECH)~\cite{rfc-draft-ietf-tls-esni-07}.

Assuming an idealistic future in which all network traffic is encrypted and
domain name information is never exposed on the wire as plaintext, packet
metadata (e.g., time, size) and destination IP addresses are the only
remaining information related to a visited website that can be seen by a
network-level observer. As a result, tracking a user's browsing history
requires the observer to infer which website is hosted on a given destination
IP address. This task is straightforward when an IP address hosts only one
domain, but becomes more challenging when an IP address hosts multiple
domains. Indeed, recent studies have shown an increasing trend of websites
being co-located on the same hosting server(s)~\cite{Shue:2007,
Hoang2020:WebCo-location}. Domains are also often hosted on multiple IP
addresses, while the dynamics of domain--IP mappings may also change over time
due to network configuration changes or DNS-based load balancing.

Given these uncertainties in reliably mapping domains to their IPs, we
investigate the extent to which accurate browsing tracking can still be
performed by network-level adversaries based solely on destination IPs. In
this work, we introduce a \emph{lightweight} website fingerprinting (WF)
technique that allows a network-level observer to identify with high accuracy
the websites a user visits based solely on IP address information,
\emph{enabling network-level browsing tracking at
scale}~\cite{Nasr2017CompressiveTA}. For instance, an adversary can
use---already collected by existing routers---IPFIX (Internet Protocol Flow
Information Export)~\cite{rfc7015} or NetFlow~\cite{NetFlow} records to easily
obtain the destination IP addresses contacted by certain users, and track
their browsing history.

For our attack, we first crawl a set of 220K websites, comprising popular and
sensitive websites selected from two website ranking lists
(\sectionref{sec:experiment_setup}). After visiting each website, we extract
the queried domains to construct a domain-based fingerprint. The corresponding
IP-based fingerprint is then obtained by continuously resolving the domains
into their IPs via active DNS measurement
(\sectionref{sec:ip_based_fingerprint}). By matching these IPs from the
generated fingerprints to the IP sequence observed from the network traffic
when browsing the targeted websites, we could successfully fingerprint 84\% of
them (\sectionref{sec:naive_evaluation}). The successful identification rate
increases to 92\% for popular websites, and 95\% for popular \emph{and}
sensitive websites.

To further enhance the discriminatory capacity of the fingerprints, we
consider the critical rendering path~\cite{Grigorik:2018} to capture the
approximate ordering structure of the domains that are contacted at different
stages while a website is being rendered in the browser
(\sectionref{sec:enhanced_fingerprint}). Our results show that the enhanced
fingerprints could allow for \emph{91\% of the tested websites} to be
successfully identified based solely on their destination
IPs(\sectionref{sec:enhanced_evaluation}).

Given the high variability of website content and domain--IP mappings across
time, we expect that once generated, a fingerprint's quality will deteriorate
quickly over time. To assess the aging behavior of the fingerprints, we
conducted a longitudinal study over a period of two months. As expected,
fingerprints become less accurate over time, but surprisingly, after two
months, they are still effective at successfully identifying about 70\% of the
tested websites (\sectionref{sec:fingerprint_stability}).

As our WF technique is based on the observation of the IPs of network
connections that fetch HTTP resources, it is necessary to evaluate the impact
of HTTP caching on the accuracy of the fingerprints. This is because cached
resources can be loaded directly from the browser's cache when visiting the
same website for a second time, resulting in the observation of fewer
connections per fingerprint. Furthermore, our attack exploits the fact that
websites often load many external resources, including third-party analytics
scripts, images, and advertisements, making their fingerprints more
distinguishable. We thus also investigated whether the removal of these
resources due to browser caching or ad blocking could help to make websites
less prone to IP-based fingerprinting
(\sectionref{sec:fingerprint_robustness}).

By analyzing the HTTP response header of the websites studied, we find that
86.1\% of web resources are cacheable, causing fewer network connections to be
observable by the adversary if these resources are loaded from the browser's
cache (\sectionref{sec:fingerprint_http_cache}). Moreover, using the Brave
browser to crawl the same set of websites, we found that the removal of
third-party analytics scripts and advertisements can impact the order in which
web resources are loaded (\sectionref{sec:fingerprint_brave}), significantly
reducing the accuracy of the enhanced fingerprints from 91\% to 76\%.
Nonetheless, employing the initially proposed WF technique in which the
critical rendering path~\cite{Grigorik:2018} is not taken into account, we
could still fingerprint 80\% of the websites even when browser caching and ad
blocking are considered.

Regardless of the high degree of website co-location and the dynamics of
domain--IP mappings, our findings show that domain name encryption alone is
not enough to protect user privacy when it comes to IP-based WF. As a step
towards mitigating this situation, we discuss potential strategies for both
website owners and hosting providers towards hindering IP-based WF and
maximizing the privacy benefits offered by domain name encryption. To the
extent possible, website owners who wish to make IP-based website
fingerprinting harder should try to (1) minimize the number of references to
resources that are not served by the primary domain of a website, and (2)
refrain from hosting their websites on static IPs that do not serve any other
websites. Hosting providers can also help by (1) increasing the number of
co-located websites per hosting IP, and (2) frequently changing the mapping
between domain names and their hosting IPs, to further obscure domain--IP
mappings, thus hindering IP-based WF attacks.
\vspace{-1.5em}

\section{Background and Motivation}
\vspace{-0.5em}

In this section, we review some background information on domain name
encryption technologies and discuss the motivation behind our study.
In particular, we highlight how our IP-based fingerprinting attack is
different from prior works, allowing network-level adversaries to effectively
mount the attack at scale.

\vspace{-0.5em}
\subsection{Domain Name Encryption}
\label{sec:domain_name_encryption}
\vspace{-0.5em}

In today's web browsing environment, there are two channels through which
domain name information is exposed on the wire: plaintext DNS
requests/responses, and the Server Name Indication (SNI) extension of TLS.

The plaintext nature of DNS not only jeopardizes user privacy, but also allows
network-level entities to interfere with user connections. For example, an
on-path attacker can inject forged DNS responses to redirect a targeted user
to malicious hosts~\cite{holdonDNS}. The domain name information exposed via
DNS packets and the SNI field has also been intensively exploited by
state-level network operators for censorship purposes~\cite{Pearce:2017:Iris,
hoang:2019:measuringI2P, iclab_SP20, Tripletcensors, GFWatch}. To cope with
these security and privacy problems, several solutions have been recently
proposed to safeguard domain name information on the wire, including
DoT~\cite{rfc7858}, DoH~\cite{rfc8484}, and
ECH~\cite{rfc-draft-ietf-tls-esni-07}.

By encrypting DNS traffic, DoT/DoH preserves the integrity and confidentiality
of DNS resolutions. Several companies (e.g., Google~\cite{googleDoH},
Cloudflare~\cite{cloudflare_DoH}) offer free DoT/DoH services to the public,
while popular web browsers including Chrome and Firefox already support
DoH~\cite{googleDoH,firefoxDoH_start}, with the latter enabling DoH (through
Cloudflare) by default for users in the US since 2019. However, the design
choice of these vendors to centralize all DNS resolutions to one trusted
resolver has raised several privacy concerns. As a result, more
privacy-centric DNS resolution schemes have also been proposed,
including DoH over Tor~\cite{cloudflareDNS-Tor, Muffett2021},
Oblivious DNS~\cite{schmitt_oblivious, Singanamalla2021}, and distributed DoH
resolution~\cite{Hoang2020:MADWeb},
to not only conceal DNS packets from on-path observers,
but also to deal with ``nosy'' recursors.

SNI has been incorporated into the TLS protocol since 2003~\cite{rfc3546} as a
workaround for name-based virtual hosting servers to co-host many websites
that support HTTPS. During the TLS handshake, the client includes the domain
name of the intended website in the SNI field in order for the server to
respond with the corresponding TLS certificate of that domain name. Until TLS
1.2, this step takes place before the actual encryption begins, leaving the
SNI field transmitted in plaintext, and exposing users to similar security and
privacy risks as discussed above.
TLS 1.3 provides an option to encrypt the SNI field, concealing the domain
name information~\cite{rfc-draft-ietf-tls-esni-06}. Since March 2020, ESNI has
been reworked into the ECH extension~\cite{rfc-draft-ietf-tls-esni-07}. In
order for ECH to function, a symmetric encryption key derived from the
server's public key has to be obtained in advance. This public key can be
obtained via an HTTPS resource record lookup. Thus, it is important to note
that ECH cannot provide any meaningful privacy benefit without the use of
DoT/DoH, and vice versa. Mozilla has supported ECH since Firefox
85~\cite{firefoxECH}.

\subsection{Website Fingerprinting}
\label{sec:wfp_background}
\vspace{-0.5em}

Website fingerprinting (WF) is a type of traffic analysis attack, employed to
construct fingerprints for a set of websites based on the traffic pattern
observed while browsing them. Depending on which metadata is visible from the
encrypted traffic, different WF techniques can be used to determine whether a
user under surveillance visited any of the monitored websites.

Numerous WF attacks targeting anonymized or obfuscated communication channels
have been proposed~\cite{Liberatore2006CCS, Wang2014USEC, Panchenko2016NDSS,
Hayes:USEC2016, Nasr:CCS18, Sirinam:CCS2018, Pulls:PETS20, Wang2020SP},
in which the actual
destination IP address is hidden by means of privacy-enhancing network
relays~\cite{Goldberg:COMPCON97, Hoang2014:Anonymous}, such as
Tor~\cite{Tor04} or the Invisible Internet Project (I2P)~\cite{Hoang2018:IMC,
petcon2009-zzz}. However, WF attacks on standard encrypted web traffic (i.e.,
HTTPS), in which no privacy-enhancing network relays are employed, have not
been comprehensively investigated, especially at the IP-address level. This is
because the domain name information previously available in several plaintext
protocols (e.g., DNS, the SNI extension of TLS, and OCSP
queries~\cite{rfc6960}) can be easily obtained from the network traffic
(\sectionref{sec:domain_name_encryption}). This information alone can already
be used for a straightforward inference of applications or web services being
visited~\cite{Hoang2017:locationPrivacy, Trevisan2016TowardsWS}. However, in
an idealistic future where domain name encryption (i.e., DoT/DoH \emph{and}
ECH) is fully deployed, visibility to any information above the IP layer will
be lost. Under these conditions, and given the high degree of web
co-location~\cite{Hoang2020:WebCo-location}, our ultimate goal is to
investigate the extent to which websites can still be fingerprinted at scale,
based solely on the IP address information of the servers being contacted.

Given the numerous WF methods introduced in the past, one may wonder
\emph{why do we even need another website fingerprinting method?} In
addition to the aforementioned reasons and pitfalls of previous WF
techniques~\cite{Juarez:CCS2014}, the rationale behind our fingerprinting
technique based solely on IP-level information stems from the increasing
deployment of domain name encryption technologies~\cite{Deccio:CoNEXT19,
Lu:2019:IMC19}. Currently, most web traffic does expose domain information, as
domain name encryption has not been fully deployed~\cite{Trevisan2020DoesDN},
and thus network-level browsing tracking at scale through DNS or SNI is much
easier. However, once this massive-scale monitoring capability
is gone due to DoT/DoH and ECH, the next best option for ISPs to continue
tracking at a similar scale will be to rely on IP addresses, which are already
collected as part of IPFIX (Internet Protocol Flow Information
Export)~\cite{rfc7015} or NetFlow~\cite{NetFlow} records by existing routers.
Although more elaborate fingerprinting schemes can certainly be conceived for
HTTPS traffic, these will require a significant deployment effort and
cost~\cite{Nasr2017CompressiveTA}, both for constructing and maintaining the
fingerprints, as well as for matching them.
\vspace{-1.5em}

\section{Threat Model}
\label{sec:threat_model}
\vspace{-0.5em}

Internet service providers (ISPs) have been increasingly harvesting user
traffic for monetization purposes, such as targeted
advertising~\cite{ISP_Forbes, ISP_wsj, Gonzalez2016UserPI}. Our threat
model considers the real-world scenario in which a local adversary (e.g., an
ISP) passively monitors users' traffic and attempts to determine whether a
particular website was visited. The adversary carries out the following steps
to create website fingerprints.

First, the adversary visits a set of websites and records all domain names
that are contacted to fetch their resources. A domain-based fingerprint for
each website is then built from this set of contacted domains.
After that, these domains are periodically resolved to their hosting IPs,
which are used to construct IP-based fingerprints. Depending on the
relationship between a domain name and its hosting IP(s), a connection to a
unique IP can be used to easily reveal which website is being visited if the
IP only hosts that particular domain name.
Finally, to conduct the WF attack, the adversary tries to match the sequence
of IP addresses found in the network trace of the monitored user with the
IP-based fingerprints constructed in the previous step to infer which website
was visited.

The effectiveness of our attack depends on two primary factors, namely, the
\emph{uniqueness} (\sectionref{sec:experiment_eval}) and \emph{stability}
(\sectionref{sec:fingerprint_stability}) of the fingerprints. It is worth
emphasizing that our model does not assume fingerprinting of obfuscated
network traffic, in which the IP address information is already hidden by
means of privacy-enhancing technologies. This class of attacks, whose goal is
to use sophisticated traffic analysis methods to fingerprint anonymized
network traffic, has been extensively investigated by prior
studies~\cite{Liberatore2006CCS, Cai2014CCS, Hayes:USEC2016,
Panchenko2016NDSS, Sirinam:CCS2018}. Our threat model requires only minimal
information collected from the network traffic, i.e., destination IPs.

We consider a browsing scenario in which one website is visited at a time,
which is particularly valid when it comes to ordinary web users on devices
with smaller screens, and is also most often the case of casual browsing
behavior (except, perhaps, the rare event of a browser restart with many
previously open tabs). Although some users may visit more than one website at
a time, they mostly interact with one tab at a time. There is also a time gap
when changing from one tab to another to open or reload a different website,
especially for users with a single screen. All these together allow an
observer to distinguish between individual website visits, as also evident by
existing techniques that can be employed to split a network trace of such
multi-tab activity into individual traces~\cite{Coull2007OnWB, Xu2018ACSAC,
Cui2019RevisitingAF}. Moreover, although many individual users may be located
behind the same NAT network, Verde et al.~\cite{Verde2014NoNU} have developed a
framework that can identify different individuals behind a large metropolitan
WiFi network based on NetFlow records. To keep our study simple, we thus
assume that the adversary already employs the aforementioned techniques to
obtain the network trace of different individuals before conducting our WF
attack.
\vspace{-1.6em}

\section{Fingerprint Construction}
\label{sec:fingerprint_construct}
\vspace{-0.8em}

Next, we explain how we construct IP-based fingerprints in more detail, from
creating the initial domain-based fingerprints to deriving the final IP-based
fingerprints. At a conceptual level, we first explore the straightforward
approach of resolving all domains loaded while visiting a website to their
hosting IPs, from which we create a set of unique IPs that can potentially be
used as the IP-based fingerprint for that website. We then take the
critical rendering path~\cite{Grigorik:2018} into account to improve the
fidelity of the fingerprints, by considering the approximate order in which
domains are loaded while the website is being rendered on the screen.

\vspace{-1em}
\subsection{Basic IP-based Fingerprint}
\label{sec:ip_based_fingerprint}
\vspace{-0.5em}

Assuming an idealistic web browsing scenario in which domain name information
can no longer be extracted from network traffic due to the full deployment of
domain name encryption, the only remaining information visible to the
adversary is packet metadata (e.g., time, size) and sequences of connections
to remote IP addresses of contacted web servers. Under these conditions, the
adversary would need to fingerprint targeted websites based primarily on this
information.
As introduced in our threat model, the adversary first visits the targeted
websites and records all domains that are contacted while browsing each
website. Each domain can then be resolved to its hosting IP address(es). As a
result, the mapping between domains and hosting IP address(es) is the
basic unit on which the adversary relies to construct IP-based fingerprints.

When browsing a website, the browser first contacts the web server to fetch
the initial resource---usually an HTML document. It then parses the HTML
document and subsequently fetches other web resources referenced. Based on
this underlying mechanism of fetching a webpage, the adversary constructs
domain-based fingerprints as follows. For a given website, its domain-based
fingerprint consists of two parts: the \emph{primary} domain, denoted as
$d_{p}$, and a set of \emph{secondary} domains, denoted as $d_{s}$. The
fingerprint then can be represented as: $d_{p}$ + $\{d_{s_1}, d_{s_2}, ...,
d_{s_n}\}$. The primary domain is the domain of the URL shown in the browser's
address bar, and typically corresponds to the server to which the first
connection is made for fetching the initial HTML document of the visited
webpage. Secondary domains may be different from the primary one and are used
for hosting other resources needed to load the webpage.

From the domain-based fingerprints constructed above, the adversary can then
obtain their corresponding IP-based fingerprints by repeatedly resolving the
domain names into their hosting IP(s). Given that a domain name may be
resolved to more than one IP, each domain in a domain-based fingerprint is
converted to a set of IP(s) with at least one IP in it. As domain-based
fingerprints are comprised of two parts, the inherent structure of IP-based
fingerprints also consists of two parts. The first part contains the IP(s) of
the primary domain name, while the second part is a set of sets of IPs
obtained by resolving the secondary domains. The fingerprint then can be
represented as:
\vspace{-1.2em}

\begin{equation}
\begin{gathered}
\{{d_{p}{ip}_{1}}, {d_{p}{ip}_{2}}, ..., {d_{p}{ip}_{n}}\} +
\{\{{d_{s_1}{ip}_{1}}, {d_{s_1}{ip}_{2}}, ..., {d_{s_1}{ip}_{n}}\}, \\
\{{d_{s_2}{ip}_{1}}, {d_{s_2}{ip}_{2}}, ..., {d_{s_2}{ip}_{n}}\}, ...,
\{{d_{s_n}{ip}_{1}}, {d_{s_n}{ip}_{2}}, ..., {d_{s_n}{ip}_{n}}\}\}
\end{gathered}
\end{equation}

To simplify the construction and matching of fingerprints, we reduce the above
fingerprint by considering the union of the sets of IP addresses of all
secondary domains (second part of the above fingerprint). The simplified
version of the IP-based fingerprint can thus be represented by just two sets
of IP addresses as:
\vspace{-1.2em}

\begin{equation}
\begin{gathered}
\{{d_{p}{ip}_{1}}, {d_{p}{ip}_{2}}, ..., {d_{p}{ip}_{n}}\} +
\{{d_{s_1}{ip}_{1}}, {d_{s_1}{ip}_{2}}, ..., {d_{s_1}{ip}_{n}}, \\
{d_{s_2}{ip}_{1}}, {d_{s_2}{ip}_{2}}, ..., {d_{s_2}{ip}_{n}}, ...,
{d_{s_n}{ip}_{1}}, {d_{s_n}{ip}_{2}}, ..., {d_{s_n}{ip}_{n}}\}
\end{gathered}
\end{equation}

Although it might seem that this simplification discards some part of the
structural information of the page, we found no significant difference in
accuracy when evaluating both fingerprint formats. Therefore, we opt to use
the latter fingerprint structure, as it is simpler and allows for faster
matching.

\vspace{-1.5em}
\subsection{Enhanced IP-based Fingerprint with Connection Bucketing}
\label{sec:enhanced_fingerprint}
\vspace{-0.5em}

When a webpage is visited, besides the initial connection to the primary
domain, multiple requests may then be issued \emph{in parallel} to fetch other
resources referenced in the initial HTML document. Once fetched, these
resources may sometimes trigger even more requests for other sub-resources
(e.g., JavaScript). The absolute order of these requests on the wire can
change from time to time, depending on many uncertain factors, such as the
performance of the upstream network provider and the underlying operating
system. For that reason, we did not consider the order in which domains are
contacted when constructing the domain-based fingerprints, and thus we gather
all secondary domains into one bucket, as described
in~\sectionref{sec:ip_based_fingerprint}. However, when viewing all these
requests as a whole, there still exists a high-level ordering relationship
that we can capture when considering the critical rendering
path~\cite{Grigorik:2018}. Specifically, there are certain render-blocking and
critical objects that always need to be loaded prior to some other objects.

The chain of events from fetching an initial HTML file to rendering the
website on screen is referred to as the \emph{critical rendering
path}~\cite{Grigorik:2018}. When visiting a website, the browser first
contacts the primary domain to fetch the initial HTML file (e.g.,
\texttt{index.html}). Next, this file is parsed to construct the DOM (Document
Object Model) tree. The browser then fetches several web resources from remote
destinations to render the webpage. Depending on the complexity of the
webpage, these resources may include HTML, JavaScript, CSS, image files, and
third-party resources, which may in turn load more sub-resources hosted on
other third-party domains~\cite{Nikiforakis2012}. According to the Internet
Archive, a typical website loads an average of 70 web resources as of this
writing~\cite{httpArchive}. When considering this critical rendering path,
there are three important events that we can use to cluster connections into
three ``buckets:'' \emph{domLoading}, \emph{domContentLoaded}, and
\emph{domComplete}.

\myparab{domLoading} is triggered when the browser has received the initial
HTML file and parses it to construct the DOM tree. As a result, multiple
parallel connections to fetch critical resources referenced by the DOM tree
are initiated right after this event is fired.

\myparab{domContentLoaded} is triggered when both the DOM and CSSOM (CSS
Object Model) are ready~\cite{Grigorik:2018}, signaling the browser to create
the render tree. The event is typically fired without waiting for style
sheets, images, and subframes to load~\cite{DOMContentLoaded}. After this
event, subsequent connections can often be observed for fetching elements such
as non-blocking style sheets, JavaScript files, images, and subframes.

\myparab{domComplete} is triggered when the website and its sub-resources have
been loaded. After this event is fired, non-essential objects can still be
downloaded in the background, leaving the critical rendering path unaffected.
For example, external JavaScript files are known to be render-blocking and are
recommended to be moved to the end of the webpage or to be included with a
\texttt{defer} attribute of the \texttt{<script>} tag~\cite{BlockingJS}.
Therefore, a small cluster of connections can often be observed after this
event is triggered.

Based on these observations about the critical rendering path, we enhance the
structure of our fingerprints (both domain-based and the corresponding
IP-based ones) to comprise the primary domain and three sets of domains
corresponding to the three events aforementioned. The enhanced domain-based
fingerprint can then be represented as:

\vspace{-1.5em}
\begin{equation}
\begin{gathered}
d_{p} + \{d_{s_1}, d_{s_2}, ...\} + \{d_{s_2}, d_{s_3}, ...\} + \{..., d_{s_{n-1}}, d_{s_n}\}
\end{gathered}
\end{equation}

Note that a domain can appear in more than one bucket if multiple objects are
fetched from that domain at different times. Accordingly, the representation
of the enhanced IP-based fingerprint follows the same structure, comprising
four sets of IP addresses: i)~the set of IP addresses of the primary domain,
and ii)~three sets of IP addresses corresponding to the three buckets above.
\vspace{-2em}

\section{Experiment Setup}
\label{sec:experiment_setup}
\vspace{-.8em}

In this section, we provide the details of how we set up and conducted our
experiments for assessing the effectiveness of IP-based website
fingerprinting. In particular, we discuss the rationale behind our test list
of websites, and the duration and location of our measurement.

\vspace{-1em}
\subsection{Selection of Test Domains}
\label{sec:test_list}
\vspace{-.8em}

From an adversarial point of view, it is desirable for an attacker to be able
to reveal as many websites as possible. It is, however, impractical to crawl
the entire Internet, given that there are more than 362.3 million domain names
registered across all top-level domains (TLDs) as of
2020~\cite{verisign.domains}. In addition, many of them are dormant or even
unwanted domains~\cite{Szurdi:usenixsecurity14} that the majority of Internet
users will never visit. As our goal is to assess the extent to which domain
name encryption would prevent the leakage of the majority of users' browsing
activities via IP-based website fingerprinting, we opt to focus on those
websites that are legitimately visited in real-world scenarios. Therefore, we
choose to use domains from the Tranco top-site ranking
list~\cite{LePochat2019}, since it has been shown to have a good overlap with
the web traffic observed by the Chrome User Experience
Report~\cite{LePochat:CSET2019}.

The research community often relies on one of the four top-site lists
(Alexa~\cite{alexa}, Majestic~\cite{majestic}, Umbrella~\cite{cisco_umbrella},
and Quantcast~\cite{quantcast}). However, studies have discovered several
issues with these lists that can negatively impact research outcomes if not
handled properly~\cite{LePochat2019, Rweyemamu2019}. To remedy the
shortcomings of these lists, the Tranco list is curated
to aggregate the four aforementioned lists, resulting in a list of more than
seven million domains. Even then, due to the dynamic nature of the
web~\cite{levene2004web, Baeza-Yates2004}, there are still domains that are
unstable, not responsive, or do not serve any web content in the Tranco list,
especially in its long tail~\cite{LePochat:CSET2019, Rweyemamu2019}.
Therefore, we select the top 100K popular domains from the Tranco list for our
study, because any ranking under 100K is not statistically significant, as
suggested by both top-list providers and previous studies~\cite{alexa,
Rweyemamu2019}.

While some websites are so common that visiting them may be considered to be a
very low privacy risk (e.g., \texttt{facebook.com}, \texttt{twitter.com}, or
\texttt{youtube.com}), the leakage of visits to more ``sensitive'' websites is
definitely an important concern. This is even more so in oppressive regions,
where browsing certain online content could be considered as a violation of
local regulations~\cite{ChinaBansVPN, ChinaBansGame, IndiaJail}. To that end,
we complement our dataset by manually choosing websites from the Alexa list
that belong to categories deemed ``sensitive,''\footnote{It is worth noting
that sensitivity can be different from site to site, depending on who, when,
and from where is visiting the site~\cite{Hoang2017:locationPrivacy}. We
intuitively choose these complementary domains based on our common sense of
what is sensitive based on those categories that are often blocked by many
Internet censors around the world~\cite{iclab_SP20}.} such as LGBT, sexuality,
gambling, medical, and religion. In total, our dataset consists of 220,743
domains, including the top 100K popular domains of the Tranco
list\footnote{The list was created on March 3rd 2020, and is available at
\url{https://tranco-list.eu/list/J2KY}.} and 126,597 domains from Alexa's
sensitive categories. Among these, there are 5,854 common domains between the
two data sources.

Although one may consider our test list as a closed-world dataset, it is
infeasible to repeatedly crawl the entire Internet, which has more than 362.3M
domains registered at the time of our experiment~\cite{verisign.domains}. It
is also unlikely that a network-level adversary is interested in
fingerprinting all websites on the Internet. However, the ability of
continuously conducting our WF attack on more than 220K domains highlights the
scalability of our method. In fact, we cover an order of magnitude
 more domains compared to
previous WF attacks against DoT/DoH traffic~\cite{bushart_padding,
Houser:2019:CoNEXT19, Siby:NDSS20}, in which the largest open-world setting
comprised fewer than 10K domains~\cite{bushart_padding}.

\vspace{-.8em}
\subsection{Measurement Duration and Location}
\label{sec:experimental_duration_location}
\vspace{-.8em}

Prior work often overlooks the temporal aspect when constructing fingerprints.
More specifically, many efforts are conducted in a one-off manner or over a
short period of time, neglecting the temporal characteristics of fingerprints
whose websites' content may evolve over time~\cite{levene2004web}. For our
work, in which we focus on IP-based fingerprints, the churn in domain--IP
mappings is also another major concern~\cite{Hoang2020:ASIACCS}. Due to the
variability of website content and hosting IPs across time, a previously
constructed fingerprint may not be valid after a certain time period.
Therefore, a longitudinal measurement study is essential to examine the
robustness of fingerprints, which in turn impacts the efficacy of their use in
WF attacks.

Over a period of 60 days (from March 5th to May 3rd, 2020), we repeatedly
crawled the 220K websites from our test list curated in
\sectionref{sec:test_list}, using the Chrome browser (desktop version 80.0),
running on Ubuntu 20.04 LTS. When visiting each website, we extract all
domains contacted to construct the fingerprint for that website using the
steps discussed in~\sectionref{sec:fingerprint_construct}. At the network
level, we capture the sequence of destination IPs contacted, to evaluate the
accuracy of our IP-based WF method (\sectionref{sec:experiment_eval}). Note
that the collection of this sequence of IPs is oblivious to the domains
extracted independently when loading each website.

Due to DNS-based load balancing, many domains, especially of popular websites,
may map to different IP addresses at different times~\cite{Hoang2020:ASIACCS}.
Therefore, once the set of domains that were contacted to render each website
is extracted, we continuously resolve them to obtain their IP addresses until
the next crawl. This best-effort approach allows us to obtain as many IPs as
possible for those domains that employ DNS-based load balancing. However, to
make sure our experiment does not saturate DNS servers (thus affecting other
legitimate users), we enforce a rate limit of at least three hours. In other
words, contacted domains of a website are only resolved again if they were not
resolved within the last three hours. The entire process for each crawl batch
takes approximately 2.5 days. As a result, we have collected a total of 24
data batches during a two-month period. To stimulate future studies in this
research domain, we make our dataset available to the research community at
\texttt{\url{https://homepage.np-tokumei.net/publication/publication_2021_popets}}.

Our measurement is conducted from a cluster of machines located in a gigabit
academic network in the US. Due to the rapid increase in the use of content
delivery networks (CDN), web content can be served from multiple servers
distributed across different locations, depending on the origin of the
request~\cite{Hoang2020:ASIACCS}. Although our dataset can be considered as
representative for web users within our geographical area, it would have
missed some IP addresses of CDN-hosted websites which can only be observed at
other locations. Nonetheless, as mentioned in~\sectionref{sec:threat_model},
the adversary in our threat model is local and also has access to the Internet
from the same network location as the monitored users (e.g., the ISP of a home
or corporate user), and will not observe any other IPs of CDN-hosted websites
either. Adversaries at different locations can always set up machines within
their network of interest and conduct the same experiment with ours to
construct a dataset of fingerprints that matches those websites browsed by
users within the network of their control.
\vspace{-1.5em}

\section{Fingerprinting Accuracy}
\label{sec:experiment_eval}
\vspace{-.5em}

Next, we evaluate the accuracy of our WF techniques using the data collected
in~\sectionref{sec:experiment_setup}. We begin with an analysis of the
information entropy that we can expect from domain-based fingerprints, and
then evaluate the accuracy of IP-based fingerprints.

\vspace{-1.2em}
\subsection{Fingerprint Entropy}
\label{sec:fingerprint_entropy}
\vspace{-.5em}

As discussed in~\sectionref{sec:threat_model}, the creation of IP-based
fingerprints is based on the domains contacted while visiting the targeted
websites. Therefore, it is important to first examine the uniqueness of these
domains, as it impacts the effectiveness of IP-based WF. This will aid us in
deciding whether a domain should be included or not as part of a fingerprint.
For example, if a certain domain is contacted when visiting every single
website, then there is no point in including it. In contrast, if a unique
domain is only contacted when visiting a particular website, it will make the
fingerprint more distinguishable. The more unique a domain is, the higher the
information entropy that can be gained~\cite{Shannon:Entropy}, resulting in a
better fingerprint.

Let $P(d)$ be the probability that a particular domain will be contacted when
visiting a given website among the targeted websites. The information entropy
in bits gained if that particular domain is contacted can be calculated using
the following formula:

\vspace{-1.3em}
\begin{equation}
\text{Information entropy} = -\log_{2} P(d)
\label{eq:Shannon_entropy}
\end{equation}

\begin{figure}[t]
\centering
\includegraphics[width=0.75\columnwidth]{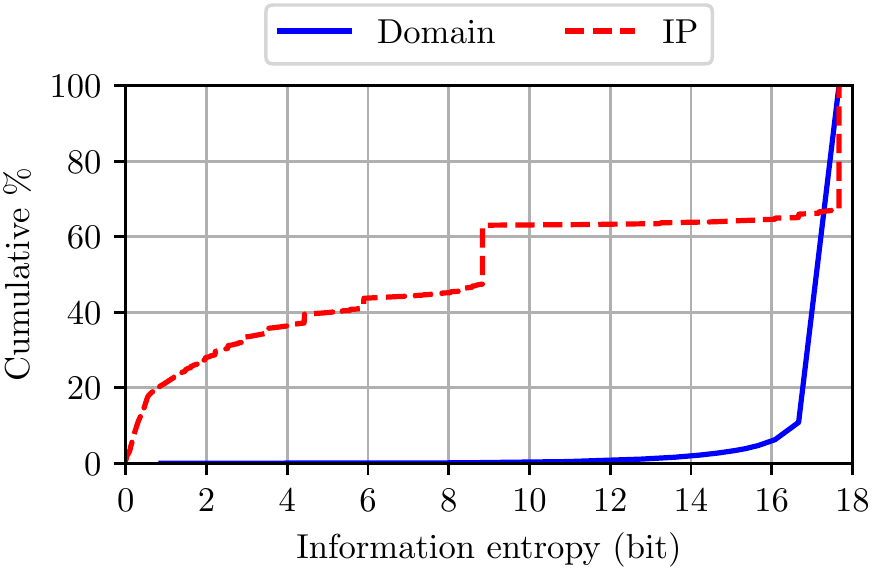}
\caption{CDF of the information entropy gained per domain/IP as a percentage
of all the unique domains/IP addresses observed.}
\label{fig:entropy}
\vspace{-1.5em}
\end{figure}

The blue (solid) line in Figure~\ref{fig:entropy} shows the entropy gained
from each domain as a percentage of the approximately 475K unique domains
observed in each crawl batch. Almost 90\% of these domains are unique to the
website from which they are referenced, yielding high information entropy
(17.7 bits). In contrast, there are a few domains from which we can only gain
a small amount of entropy. This result aligns well with the study of
Greschbach et al.~\cite{Greschbach2017TheEO} in which the Alexa top-site list
was crawled and for 96.8\% of the websites there exists at least one domain
that is unique only to these websites.

Table~\ref{tab:lowest_entropy_domains} lists the top-ten domains that provide
the least information entropy. Eight of them belong to Google and two belong
to Facebook. These domains are commonly included in many websites, thus only
contributing a small amount of entropy.
For instance, \texttt{www.google-analytics.com} is included in more than half
of the websites. However, we opt to keep them as part of our fingerprints, as
most of them still provide more than one bit of information, helping to
differentiate between websites that reference these external Google/Facebook
services and those that do not.

Based on the entropy for each domain computed above, we then calculate the
entropy gained per IP address, since our ultimate goal is to perform WF at the
IP level. Given an observed IP, there are two possibilities regarding the
domain(s) it may correspond to. First, the IP may be associated with only a
single domain, in which case its entropy can be deduced directly from the
domain's entropy. Second, the IP may co-host multiple domains. In this case,
the IP's entropy is calculated by taking the average of the entropy values of
all domains (that have been observed to be) hosted on it. Note that
calculating the entropy using both average and median gives us similar results
because most co-hosted domains on the same IP addresses often provide a
similar amount of information entropy. We thus choose the former one.

Considering these two possibilities, we then calculate the information entropy
of the 340K IPs observed from our continuous DNS measurement in each crawl
batch. As indicated by the red (dashed) line of Figure~\ref{fig:entropy}, 50\%
of the IPs provide at least 9 bits of information entropy, while there is a
group of more than 30\% of the IPs that provide a high amount of information
entropy (17.7 bits), which correspond to IPs hosting only a single domain.

\begin{table}[t]
  \footnotesize
  \centering
  \caption{Top-ten domains that yield the lowest entropy.}
  \begin{tabular}{lrc}
  \toprule
  \textbf{Domain name}                  & \textbf{\# Websites} & \textbf{Entropy} \\ [0.5ex]
  \midrule
  \texttt{www.google-analytics.com}     & 114K (55\%)          & 0.87 \\
  \texttt{fonts.gstatic.com}            & 102K (49\%)          & 1.03 \\
  \texttt{fonts.googleapis.com}         & 102K (49\%)          & 1.04 \\
  \texttt{www.google.com}               &  76K (37\%)          & 1.44 \\
  \texttt{stats.g.doubleclick.net}      &  72K (35\%)          & 1.53 \\
  \texttt{www.googletagmanager.com}     &  64K (31\%)          & 1.71 \\
  \texttt{www.facebook.com}             &  53K (25\%)          & 1.97 \\
  \texttt{connect.facebook.net}         &  53K (25\%)          & 1.98 \\
  \texttt{googleads.g.doubleclick.net}  &  49K (24\%)          & 2.09 \\
  \texttt{ajax.googleapis.com}          &  34K (16\%)          & 2.62 \\
  \bottomrule
  \end{tabular}
  \label{tab:lowest_entropy_domains}
  \vspace{-1.5em}
\end{table}

\vspace{-1.2em}
\subsection{Primary Domain to IP Matching}
\label{sec:primary_domain_evaluation}
\vspace{-.5em}

\begin{table*}[t]
\normalsize
\centering
\caption{Percentage of successfully identified websites using
i)~naive primary domain to IP matching (\sectionref{sec:primary_domain_evaluation}),
ii)~basic fingerprinting (\sectionref{sec:ip_based_fingerprint}, \sectionref{sec:naive_evaluation}), and
iii)~enhanced fingerprinting with connection bucketing (\sectionref{sec:enhanced_fingerprint}, \sectionref{sec:enhanced_evaluation}).}
\begin{tabular}{lrrrr}
\toprule
\textbf{Website type}  & \textbf{Total} & \textbf{Primary Domain}
& \textbf{IP-based Fingerprinting} & \textbf{Connection Bucketing}  \\ [0.5ex]
\midrule
All websites crawled   & 208,191 & 107,455 (52\%) & 174,662 (84\%) & \textbf{189,527 (91\%)} \\
Popular websites       &  93,661 &  58,989 (63\%) &  86,147 (92\%) &  90,231 (96\%) \\
Sensitive websites     & 120,293 &  51,538 (43\%) &  93,988 (78\%) & 104,983 (87\%) \\
Sensitive and popular  &   5,763 &   3,072 (53\%) &   5,473 (95\%) &   5,687 (99\%) \\
\bottomrule
\end{tabular}
\label{tab:fingerprinting_result_breakdown}
\vspace{-1em}
\end{table*}

Before evaluating our WF techniques, we first investigate whether WF is needed
at all. Prior studies have shown that a significant fraction of websites have
a one-to-one mapping between primary domains and their hosting
IP(s)~\cite{Hoang2020:ASIACCS}. This first connection can be distinguished
from the subsequent requests since there is usually a noticeable time gap
during which the browser needs to contact the primary domain to download the
initial HTML file, parses it, and constructs the DOM tree before multiple
subsequent connections are initiated to fetch referenced resources. As a
result, it is straightforward for an adversary to target this very first
connection to infer which website is being visited.

More specifically, when a domain is hosted on one IP or multiple IPs without
sharing its hosting server(s) with any other domains, it is easy to infer the
domain from the IP(s) of its hosting server(s). We analyzed the DNS records of
all primary domains in our dataset to quantify the fraction of websites that
can be fingerprinted by just targeting their primary IP(s).
We find that 52\% of the websites studied have their primary domain hosted on
their own IP(s), while the remaining 48\% are co-hosted on a server with at
least another website. This result means that an adversary can already infer
52\% of the targeted websites based solely on the IP address of the very first
connection to the primary domain, without having to consider secondary
connections. The third column of
Table~\ref{tab:fingerprinting_result_breakdown} shows the breakdown of these
websites in terms of their popularity and sensitivity. Note that the total
number of websites shown here is lower than the total number of test websites
selected in~\sectionref{sec:experiment_setup} due to some unresponsive
websites when conducting our experiment.

\vspace{-1.2em}
\subsection{Basic IP-based Website Fingerprinting}
\label{sec:naive_evaluation}
\vspace{-.5em}

To fingerprint the remaining 48\% of websites whose primary domains are
co-hosted, an adversary would need to analyze the second part of their
IP-based fingerprint that captures the IP addresses of secondary domains.

Going back to the way we build our IP-based fingerprints
in~\sectionref{sec:ip_based_fingerprint}, the basic IP-based fingerprint has
two parts. The first part consists of the primary domain's IP(s), and the
second part comprises a set of IPs obtained by resolving all secondary
domains. Given a sequence of IPs $[{ip}_0, {ip}_1, {ip}_2, ..., {ip}_{n}]$
observed from a network trace, we first scan ${ip}_0$ against the primary part
of all IP-based fingerprints, which are created by repeated active DNS
measurements (\sectionref{sec:experimental_duration_location}). If ${ip}_0$ is
found among the primary IPs of a given fingerprint, the fingerprint is added
to a pool of candidates. We then compare the \emph{subset} $\{{ip}_1, {ip}_2,
..., {ip}_{n}\}$ with the secondary part of each candidate fingerprint. For
each matching IP, we add the entropy provided by that IP to the total amount
of entropy gained for that particular candidate fingerprint. Finally, we
choose the fingerprint with the highest total entropy to predict the website
visited.

Using this IP-based WF method we obtained an increased matching rate of
84\%---that is, 84\% of the websites in our data set were identified with
100\% accuracy. The breakdown of the fingerprinted websites is shown in the
fourth column of Table~\ref{tab:fingerprinting_result_breakdown}. Among these
fingerprinted websites, we could precisely match 92\% of the popular websites
and 78\% of the sensitive websites. More worrisome is the fact that 95\% of
sensitive \emph{and} popular websites can be fingerprinted.

\vspace{-1.2em}
\subsection{Enhanced Website Fingerprinting with Connection Bucketing}
\label{sec:enhanced_evaluation}
\vspace{-.8em}

We next evaluate the effectiveness of the enhanced
WF~(\sectionref{sec:enhanced_fingerprint}),
in which we take the critical
rendering path into consideration to cluster IPs into three buckets. Similarly
to the basic fingerprints (\sectionref{sec:naive_evaluation}), given a
sequence of IPs $[{ip}_0, {ip}_1, {ip}_2, ..., {ip}_{n}]$, we first scan
${ip}_0$ against all fingerprints to create a pool of candidate fingerprints.
For the subsequence $[{ip}_1, {ip}_2, ..., {ip}_{n}]$, our goal is to split it
into three buckets of connections that can potentially be matched with the
three buckets of IPs in the IP-based fingerprints. Based on the time of each
connection initiation captured at the network level
(\sectionref{sec:experimental_duration_location}), we use \emph{k}-means
clustering~\cite{MacQueen1967SomeMF, Forgy1965ClusterAO, Lloyd1982LeastSQ} to
split them into three sets of IPs.

For every candidate fingerprint, we intersect each bucket of IPs in the
fingerprint to the corresponding bucket of IPs captured from the network
trace. Then, for each matching IP, we add its entropy to the total amount of
entropy gained for that particular fingerprint. Finally, we choose the
fingerprint with the highest entropy to predict the visited website.

Using this approach, the accuracy rate can be improved to 91\%. The breakdown
of fingerprinted websites is shown in the last column of
Table~\ref{tab:fingerprinting_result_breakdown}. For the popular and the
sensitive websites, we obtain an accuracy rate of 96\% and 87\%, respectively.
However, a more alarming result is that 99\% of sensitive \emph{and} popular
websites can be precisely fingerprinted, posing a severe privacy risk to their
visitors.

\begin{figure}[t]
\centering
\includegraphics[width=0.75\columnwidth]{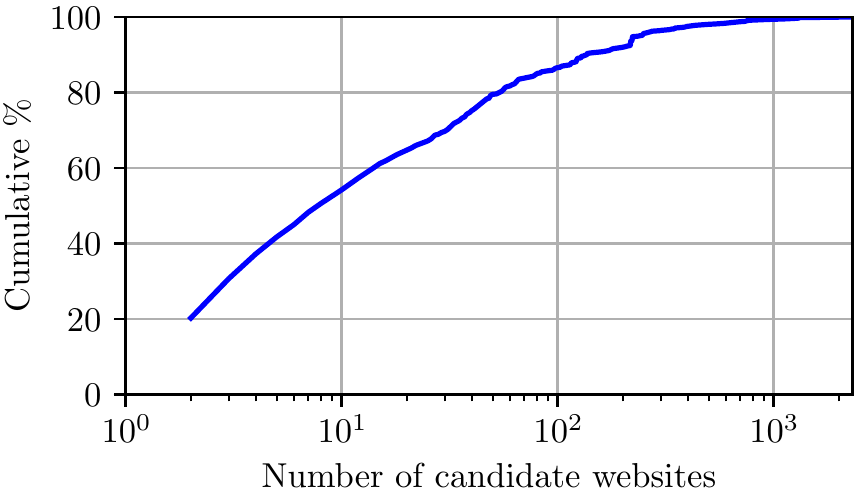}
\caption{CDF of candidate websites per fingerprint for the remaining 9\%
  (18,664) of websites that could not be matched.}
\label{fig:no_exact_match_websites_colocation}
\vspace{-1em}
\end{figure}

We now look into the remaining 9\% (18,664) of websites for which we could not
find an exact match. As shown in
Figure~\ref{fig:no_exact_match_websites_colocation}, 20\% of these websites
have only two matching candidates, while about 50\% of them have only up to
ten matching candidates. By manually examining some of these fingerprints, we
found many cases in which the matching candidate domains actually point to the
\emph{same} website (but without redirecting to the same domain). These are
mostly owned by organizations who have registered the same name under
different TLDs (e.g., \texttt{bayer.de} vs. \texttt{bayer.com}), or variations
of the domain (e.g., \texttt{christianrock.net} vs.
\texttt{christianhardrock.net}) to protect their brand against domain
squatting~\cite{Szurdi:usenixsecurity14}. A more determined adversary could
invest the effort to implement more advanced techniques for identifying such
duplicate websites. For instance, string similarity can be used to cluster
similar domains, while image similarity can be used to group websites with
similar screenshots of the start page.

\vspace{-1.5em}

\section{Fingerprint Stability}
\label{sec:fingerprint_stability}
\vspace{-.8em}

As mentioned in our threat model, the efficacy of a WF attack also depends on
the stability of fingerprints over time. There are two primary reasons why a
website fingerprint may go stale. First, the website may change over time with
existing elements removed and new elements added~\cite{levene2004web}. Second,
the mapping between a domain and its hosting IP(s) may also
change~\cite{Hoang2020:ASIACCS}. Consequently, a previously constructed
fingerprint may no longer be valid after a certain time period.

Since our IP-based fingerprints are constructed based on domains contacted
while browsing the targeted websites, we first examine the extent to which
these websites are stable in terms of the domains that they reference. We
introduce a \emph{difference} metric to quantify the change in this set of
domains for a given website over time as follows. Let $D_{t_0}$ and $D_{t_1}$
be the sets of contacted domains observed when browsing a website at time
$t_0$ and $t_1$, respectively, the \emph{difference degree} for this website
is calculated as:

\vspace{-1.1em}
\begin{equation}
\text{Difference degree} = \frac{(D_{t_0} \cup D_{t_1})-(D_{t_0}\cap D_{t_1})}{ D_{t_0} \cup D_{t_1} }
\end{equation}

Based on this definition, we consider a website as \emph{stable} during a
period $t_0 \rightarrow t_1$ when the set of domains observed at time $t_1$
have not changed compared to those previously observed at time $t_0$, yielding
a \emph{difference degree} of $0$. In contrast, a website is considered as
\emph{unstable} when its difference degree is $1$, meaning that all domains
observed at time $t_1$ are different from those previously seen at $t_0$.

\begin{figure}[t]
\centering
\includegraphics[width=0.75\columnwidth]{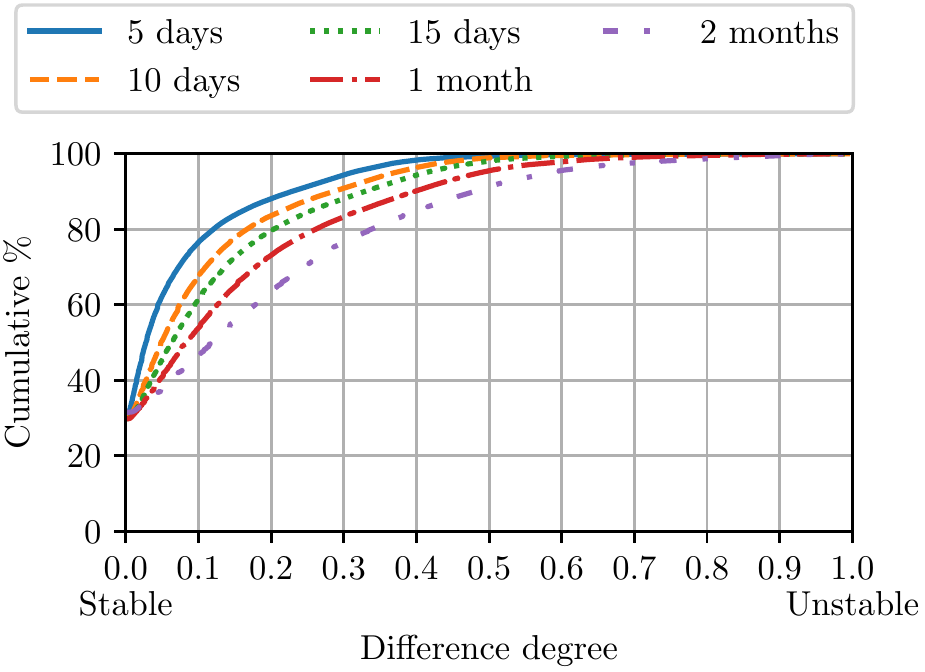}
\caption{CDF of the stability of domain names loaded in each website as a
 percentage of all websites studied.}
\label{fig:domain_stability}
\vspace{-1em}
\end{figure}

Figure~\ref{fig:domain_stability} shows the stability of the websites studied
in terms of the domains contacted while visiting them. About 30\% of the
websites contact the exact same set of domains to download web resources for
the whole two-month period of our study. Within a five-day period, 80\% of the
websites are still almost completely stable, with a difference degree lower
than 0.1, while this percentage decreases to 50\% over the two-month period.
Understandably, almost half of the websites we study are the most popular on
the Internet. Hence, it is expected that their content will be changed or
updated on a regular basis. However, even after two months, almost 80\% of the
websites are still stable, with a difference degree lower than 0.3, meaning
that 70\% of observed domains are still being used to host web resources
needed to render these websites.

This is a favorable result for the adversary, as it shows that domains
are an effective and consistent feature. The result particularly implies that
the adversary does not need to keep crawling all websites repeatedly to
construct domain-based fingerprints. Based on the results of
Figure~\ref{fig:domain_stability}, the adversary perhaps can divide websites
into two groups comprising stable and less stable websites. For instance, the
stable group consists of 80\% of websites with a difference degree lower than
0.2 after ten days, while the less stable group consists of the remaining 20\%
of websites. Then, the adversary would only need to re-crawl the less stable
ones every ten days to keep their domain-based fingerprints fresh, instead of
all websites.

\begin{figure}[t]
\centering
\includegraphics[width=0.75\columnwidth]{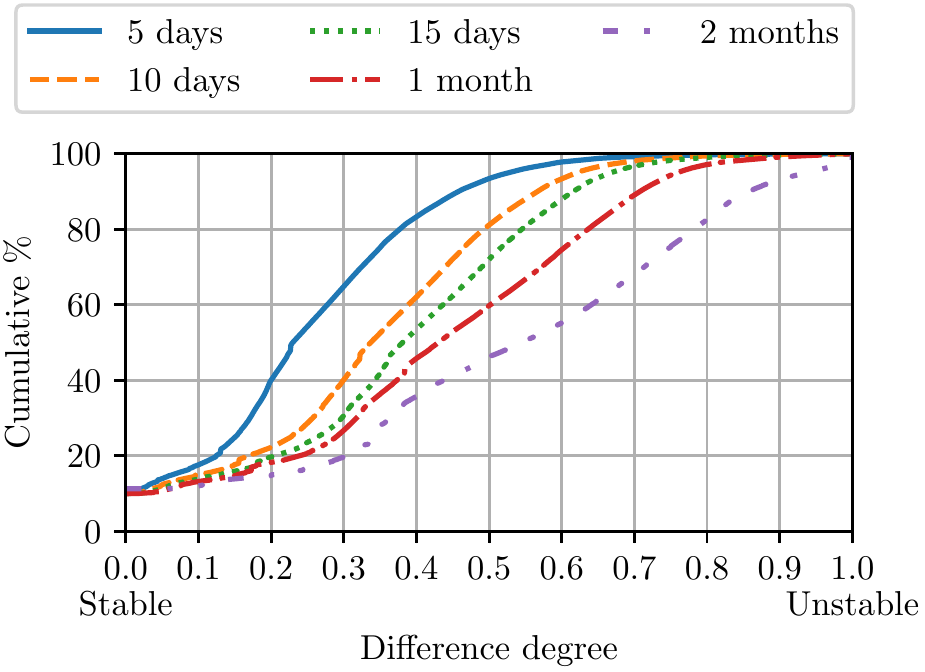}
\caption{CDF of the stability of IP addresses in each IP-based fingerprint as
a percentage of all websites studied.}
\label{fig:ip_stability}
\vspace{-1em}
\end{figure}

However, in our threat model, what can be actually observed by the adversary
is only IP addresses. We thus apply the same difference formula to quantify
the stability of IP-based fingerprints. Unlike domain-based fingerprints,
IP-based fingerprints become stale faster, as shown in
Figure~\ref{fig:ip_stability}. We find that only 10\% of the IP-based
fingerprints contain the same set of IPs over the course of two months. After
ten days, 60\% of the fingerprints have more than 30\% of their IPs changed.
After two months, half of the IPs have changed in more than 50\% of the
fingerprints.

\begin{figure}[t]
    \centering
    \includegraphics[width=0.75\columnwidth]{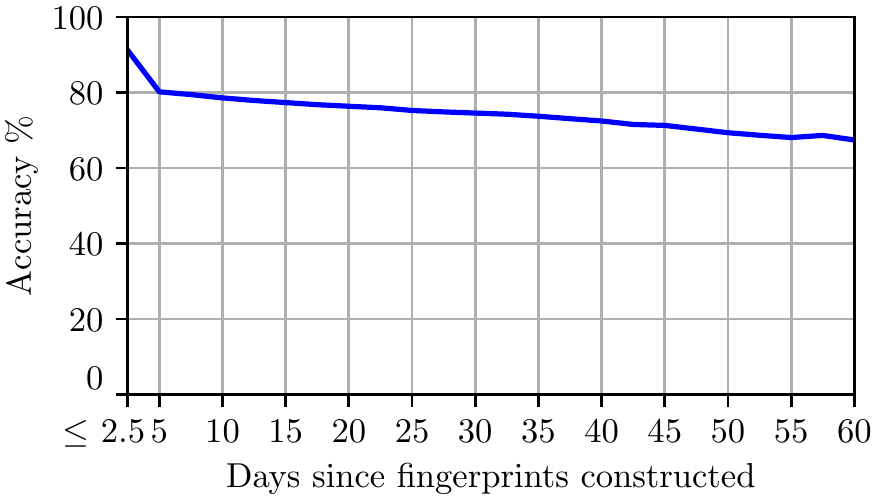}
    \caption{Fingerprint stability over time. Even after two months, 70\% of
    the tested websites can still be fingerprinted.}
    \label{fig:fingerprint_stability}
    \vspace{-1em}
    \end{figure}

Given these results, we investigate how the instability of the IP-based
fingerprints impacts the accuracy of our WF attack. We consider an attack
scenario in which the adversary uses fingerprints constructed in the past to
track the users' browsing activities at a future time.
Figure~\ref{fig:fingerprint_stability} shows the accuracy (i.e., the
percentage of successfully identified websites) of our enhanced WF approach
over the course of two months. Within 2.5 days\footnote{The lowest time
granularity is 2.5 days because each crawl batch in our dataset requires this
amount of time to be collected, as discussed
in~\sectionref{sec:experiment_setup}.} after their generation, our
fingerprints consistently yield a high accuracy of 91\%. Over the course of
two months, we can see a gradual decrease in the accuracy. However, this
decrease is quite modest, as after five to ten days since their construction
the fingerprints can still be used to accurately identify about 80\% of the
websites. This number only decreases to about 70\% after two months.

Although IP-based fingerprints go stale faster compared to their domain-based
fingerprints, those IP addresses that change frequently mostly correspond to
secondary domains, and only a small fraction corresponds to primary domains
(see Appendix~\ref{sec:Domain_ip_dynamics} for details). The vast majority of
primary domains are hosted on mostly static IP addresses for the whole period
of our study. As a result, the persistently stable IP addresses of these
primary domains in the IP-based fingerprints is the reason why our IP-based
fingerprints are still effective at revealing 70\% of the targeted websites
even though a large number of IP-based fingerprints have changed significantly
after two months, as indicated in Figure~\ref{fig:ip_stability}.

The above finding means that the adversary can intelligently split domains
into two groups, based on previously observed data. The first group consists
of domains whose IPs are dynamic, while the second group contains domains
whose IPs remain static over a configurable amount of time. The adversary then
only needs to periodically perform DNS lookups for the first group after a
desired amount of time has passed, depending on the network overhead and
resources the adversary can sustain for conducting the attack.
\vspace{-1.5em}

\section{Fingerprint Robustness}
\label{sec:fingerprint_robustness}
\vspace{-.8em}

We next examine the impact of HTTP caching on the effectiveness of our WF
since resources are often cached by web browsers to improve websites'
performance. In addition, our WF also exploits the fact that websites often
load external resources, including images, style sheets, fonts, and even
``unwanted'' third-party analytics scripts, advertisements, and
trackers~\cite{Nikiforakis2012}, which result in a sequence of connections to
several servers with different IPs, making the fingerprints more unique. Thus,
we also investigate whether blocking these unnecessary resources would help
make websites less distinguishable, thus reducing their fingerprintability.

\subsection{Impact of HTTP Caching on Website Fingerprinting Accuracy}
\label{sec:fingerprint_http_cache}
\vspace{-.8em}

\begin{figure}[t]
    \centering
    \includegraphics[width=0.75\columnwidth]{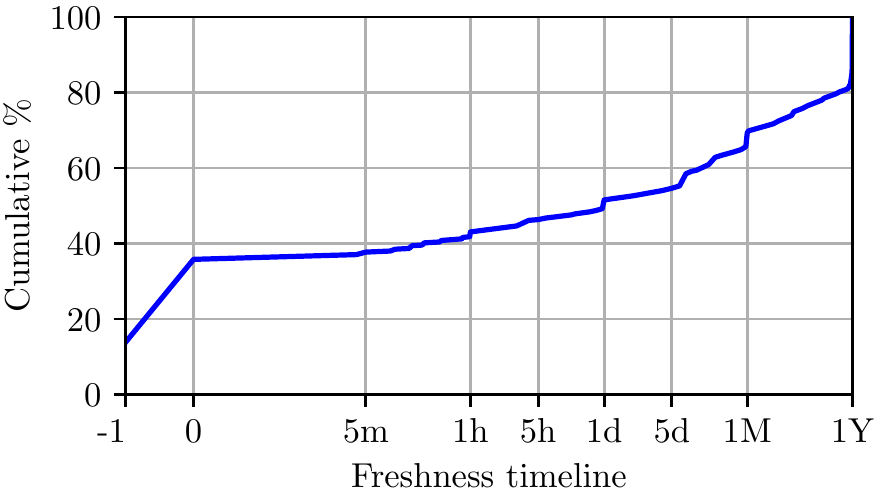}
    \caption{CDF of the freshness timeline of HTTP resources.\\(m: minute, h:
    hour, d: day, M: month, Y: year)}
    \label{fig:freshness_timeline_cdf}
\end{figure}

When a website is revisited, cached resources can be served from the local
cache without the browser fetching them again from their origins. Since our
attack is based solely on the observation of the IP of connections to remote
destinations, we are interested in examining the fraction of cacheable
resources and the extent to which HTTP caching impacts the effectiveness of
our WF.

Analyzing the response header of 21.3M objects observed while crawling the
tested websites, we find that 86.1\% of them are cacheable. In other words,
these HTTP resources can be stored and served from the local cache without
being downloaded again from the remote servers when being revisited.

Utilizing the cache-control information in the HTTP response header, we
compute the freshness timeline for each resource. The freshness timeline is
the amount of time during which the browser can store and serve resources from
its cache without downloading them again from their original servers.
Figure~\ref{fig:freshness_timeline_cdf} shows the distribution of the
freshness timeline of 21.3M objects. The value ``-1'' denotes uncachable
resources (13.9\%) that must be downloaded again from their origin if
revisited, while ``0'' indicates cacheable resources (21.9\%) that always need
to be revalidated with their origin. In other words, these two types of
resources will always cause a network connection to their original servers if
revisited. On the other hand, the remaining 64.2\% of resources can be loaded
directly from the local cache without making any network connections.

Next, we evaluate the impact of cacheable resources on our attack accuracy by
excluding IPs on which cacheable resources are hosted. We use the basic
fingerprinting method here for our evaluation
(\sectionref{sec:ip_based_fingerprint}) instead of the enhanced one
(\sectionref{sec:enhanced_fingerprint}), because revisited resources may not
be freshly loaded in the order of the critical rendering path as in the first
visit.

\begin{figure}[t]
    \centering
    \includegraphics[width=0.75\columnwidth]{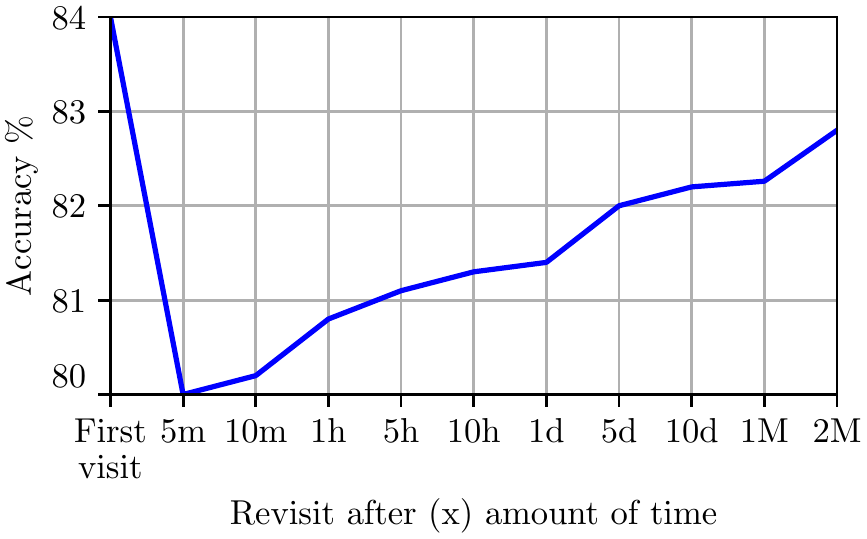}
    \caption{The accuracy of IP-based fingerprinting attack when excluding
    destinations where cacheable resources are hosted. (m: minute, h: hour, d:
    day, M: month)}
    \label{fig:fingerprint_accuracy_with_cache}
    \end{figure}

Although many web resources are cached, we could still obtain a high accuracy.
As shown in Figure~\ref{fig:fingerprint_accuracy_with_cache}, even when
websites are revisited after only five minutes, meaning that the majority of
resources can be served from the local cache, an accuracy of 80\% can still be
obtained---a decrease of just 4\% (from 84\%) compared to when websites are
visited for the first time. If websites are revisited after one hour, one day,
or one month, our basic WF attack can obtain a gradually increased accuracy of
80.8\%, 81.4\%, and 82.3\%, respectively.

There are two primary reasons why our attack is still highly effective
although the majority of resources are cacheable. First, excluded IPs often
host long-term cacheable shared resources, such as fonts and JavaScript code,
which contribute only a small amount of entropy to the fingerprint if
included. Second, for cacheable resources hosted on IPs with high entropy, not
all resources have the same freshness timeline. In fact, we find that half of
the origins that host resources cacheable for more than one hour also serve
another resource with a freshness timeline shorter than five minutes, causing
at least one network connection to the original server if revisited.

\subsection{Fingerprinting Under Ad Blocking}
\label{sec:fingerprint_brave}
\vspace{-0.8em}

Due to the prevalence of ads and analytics scripts that harvest users'
information~\cite{Razaghpanah2018AppsTP}, many advertisement and tracker
blocking tools have been developed to protect user privacy. Of these tools,
the Brave browser has stood out to be one of the best browsers for user
privacy on the Clearnet to date~\cite{Leith2020WebBP}.

Therefore, we opt to use the Brave browser for investigating the impact of ad
blocking on IP-based WF. During the last four batches of our data collection
process (i.e., ten days), at the same time with crawling the test websites
(without blocking ads and trackers), we instrumented the Brave browser
(desktop version 1.6.30) to crawl these websites a second time. While the
Brave browser is loading each website, we also capture (1) the set of domains
contacted to fetch web resources for rendering the website, and (2) the
sequence of IPs contacted to fetch web resources.

Using the same fingerprinting techniques as
in~\sectionref{sec:experiment_eval}, we then match the sequences of IP
addresses observed while browsing with the Brave browser to our IP-based
fingerprints. As expected, the fingerprinting accuracy rate decreases from
91\% to 76\% when using the enhanced fingerprints. Since our enhanced
fingerprinting approach (\sectionref{sec:enhanced_fingerprint}) primarily
relies on the ordering structure in which web resources are loaded, the
partial removal of these resources by the Brave browser has impacted the
effectiveness of this approach.

However, when employing the initial (basic) fingerprinting approach
(\sectionref{sec:ip_based_fingerprint}), in which the ordering structure of
loaded resources is not considered, we could still obtain an accuracy rate of
80\%. This is due to the removal of external resources (e.g., analytics
scripts, tracking images) whose information entropy is not significant. We
provide a further analysis of the filtered domain names in
Appendix~\ref{sec:brave_filtered_domains}.

In practice, the use of ad blocking (if any) by a client may not be explicit
to an adversary since IPs are the only information that can be observed. To
detect the use of ad blocking at the client side, the adversary can obtain
up-to-date IP blocklists of ads/trackers from well-maintained sources (e.g.,
FireHOL~\cite{firehol}) to examine if there are connections to IPs of servers
where those ads/trackers are hosted. Then, the adversary can decide which
matching mode to employ for obtaining a higher fingerprinting accuracy. Note
that the data collection procedure does not change regardless of the mode.
\vspace{-1.5em}

\section{Countermeasures}
\label{sec:countermeasures}
\vspace{-.8em}

We next discuss potential directions for website owners and hosting providers
toward making IP-based WF more challenging, thus maximizing the privacy
benefits provided by domain name encryption.

\vspace{-.8em}
\subsection{Website Owners}
\label{sec:website_owners}
\vspace{-.8em}

Our WF exploits the fact that websites typically load resources from
multiple servers. From the viewpoint of a network observer, this makes their
fingerprints more distinguishable. External resources such as ads and
tracking scripts served from third-party domains may sometimes fetch even more
``unwanted'' objects from other third-party domains without the knowledge of
the website owner~\cite{Nikiforakis2012}. As shown
in~\sectionref{sec:fingerprint_robustness}, blocking these objects hinders (to
some extent) the fingerprintability of a website. Owners who wish to provide
increased privacy to their visitors can thus minimize the inclusion of
third-party resources. On the other hand, privacy-conscious users can use ad
and tracker blocking tools to make their browsing activities harder to
fingerprint.

Another reason for contacting domains that are different from the primary
domain is the web design strategy known as \emph{domain
sharding}~\cite{Domain_sharding}. Since traditional web browsers limit the
number of concurrent connections per remote server according to the HTTP/1.1
specification~\cite{rfc2616}, website owners often host resources on different
domains as a workaround to improve the page load time by parallelizing
connections to multiple servers. However, the introduction of HTTP/2 makes
this strategy irrelevant.

Among the many new features of HTTP/2, \emph{server push} and \emph{request
multiplexing} play an important role in improving page load
time~\cite{Wang:NSDI14}. By eliminating round-trip requests, the
server can preemptively push referenced resources to reduce latency. With
multiplexing, resource requests can be sent in parallel through a single TCP
connection. To gain any performance benefits offered by these new mechanisms,
it is recommended to co-host web resources on the same
server~\cite{Wang:NSDI14}. From the perspective of IP-based fingerprinting,
this is a welcome change that will aid in reducing the fingerprintability of
websites, as a network-level observer will now see only one connection stream
to a single remote IP address.

\vspace{-.8em}
\subsection{Hosting providers}
\label{sec:hosting_providers}
\vspace{-.8em}

Even with HTTP/2 and all resources served from the same domain, if a website
is exclusively hosted alone on the same IP(s), it can still be trivially
fingerprinted. Hosting providers can aid in hindering IP-based WF by
maximizing two factors: the co-location degree of websites and the
dynamics of domain--IP mappings.

We have shown that websites that are not co-hosted with other websites are the
most vulnerable to our attack due to the one-to-one mapping between their
domain and hosting IP address(es). When a website is co-hosted with many other
websites, it becomes more challenging to fingerprint---assuming their owners
have taken the steps discussed in~\sectionref{sec:website_owners}. Otherwise,
despite a relatively high level of co-location of more than 1K websites, we
could still successfully fingerprint them because their fingerprints are
unique enough to differentiate a given website from the rest of the co-hosted
websites, as shown in Appendix~\ref{sec:secondary_analyses}.

In addition to increasing the co-location degree, hosting providers can also
maximize the dynamics of domain--IP mappings to hinder WF further. By
analyzing the dynamics of mappings between domains and IPs throughout the
whole period of our study, we find that it is feasible to increase the
dynamics of domain--IP mappings from the perspective of hosting providers.
However, we only observe this behavior for a small number of primary and
secondary domains, whereas almost 80\% of the studied websites have their
primary domains hosted on static IPs, allowing network-level observers to
easily fingerprint them (see Appendix~\ref{sec:Domain_ip_dynamics} for
details).

\vspace{-1.5em}

\section{Limitations}
\label{sec:limitations}
\vspace{-.8em}

In this section, we discuss the limitations and the rationale behind the
experimental design of our study, especially in relation to the critical
evaluation conducted by Juarez et al.~\cite{Juarez:CCS2014}.

Most prior WF studies are often criticized for only considering a very small
number of websites in a closed-world setting~\cite{TorCritique,
Juarez:CCS2014}. However, it would be infeasible to crawl the entire Internet
of more than 362.3 million domains registered across all
TLDs~\cite{verisign.domains}, many of which are dormant domains that most
users would never visit~\cite{Szurdi:usenixsecurity14}. We thus use the Tranco
top-site ranking list~\cite{LePochat2019} to focus on those websites that are
likely visited by most legitimate Internet users in real-world
scenarios~\cite{LePochat:CSET2019}. Moreover, our test list is not only
curated from the most 100K popular domains, but is also complemented by more
than 100K less popular but sensitive domains~\sectionref{sec:test_list}.

We believe that this is a reasonable trade-off for the breadth of coverage,
which yields a manageable yet representative set of test domains, allowing us
to conduct our experiment in a longitudinal fashion to shed light on the aging
behavior of fingerprints (\sectionref{sec:fingerprint_stability})---an
important factor that is often not considered by prior studies. Regardless of
the test list size, our WF attack was conducted in a closed-world setting. In
a truly open-world setting, as the number of websites increases, the proposed
WF technique may become more error-prone~\cite{Panchenko2011WPES, TorCritique,
Juarez:CCS2014}.

Another criticized assumption often made by previous studies is that the
adversary can collect data under the same conditions (e.g., network
connection, web browser, website localization) as the
victim~\cite{Juarez:CCS2014}. This is a valid criticism, especially when it
comes to WF attacks on Tor traffic, because visiting the same domain via
different Tor paths (exit nodes) may result in different localized versions of
the website. The adversary in our threat model, however, is a local attacker
(e.g., ISPs, corporate network administrators) who is in the same network with
the victim. Therefore, it is straightforward for the adversary to set up an
environment that is similar to that of the victim. Specifically, the
availability of several OS and device fingerprinting tools based on the
different implementations of the TCP/IP stack~\cite{Beverly2004:PAM,
Shamsi2016HershelSO}, together with well-known ``home-phoning'' traffic of
different web browsers~\cite{Leith2020WebBP}, can assist the adversary in
filtering background noise and resembling a similar browsing environment with
the victim.

To keep our experiment manageable, we opt to use the Chrome browser for data
collection because it is the most popular at the time of this study, occupying
about 65\% of the browser market share~\cite{statcounter}. Although the
cross-browser fingerprinting result in~\sectionref{sec:fingerprint_brave} has
showed that our basic WF technique can still achieve an accuracy rate of 80\%,
we acknowledge that this accuracy could be impacted in more complex scenarios
if different extensions and preferences are configured in the browser.

Our dataset is created by visiting the start page of the test websites without
going into any subpages or interacting with them. We thus may have missed some
characteristics of individual pages that could only be captured if some user
interaction was involved (e.g., logging in). However, similarly to DNS-based
monitoring, our attack model does not aim to distinguish between different
links, pages, or events under the same website, which has been studied
previously~\cite{Miller:PETS14, Mariano:ARES19}. Instead, the primary goal of
the proposed WF technique is to determine if a given website was visited.

When conducting our attack, the resources of each website are considered
independently for that given website while there could be cases in which more
than one website is visited from the same browser, resulting in same resources
(e.g., font, CSS, or JavaScript files) being shared among the websites. We
have shown in the analysis of our fingerprint robustness
(\sectionref{sec:fingerprint_robustness}) that excluding these resources from
the fingerprints due to browser caching or ad blocking can significantly
impact the effectiveness of our enhanced WF technique
(\sectionref{sec:enhanced_fingerprint}). Nonetheless, their removal does not
completely thwart our attack, as we could still identify 80\% of the websites
studied using the basic WF technique (\sectionref{sec:ip_based_fingerprint}).
This is because these shared resources are often hosted on common IP addresses
that contribute only a small amount of entropy to the fingerprints when
included.

Finally, the proposed privacy-enhancing countermeasure of increasing website
co-location can lead to another privacy concern related to hosting
centralization~\cite{Libert2019GoodNF}. While this is a valid concern, this
suggestion is (1) for hosting providers who are already chosen by the website
owners to host their websites, and (2) based on the already centralized nature
of the web, which has been an increasing trend for the last
decade~\cite{Shue:2007, Hoang2020:WebCo-location}. Note that the adversary in
our threat model (\sectionref{sec:threat_model}) corresponds to local
attackers, such as ISPs and corporate network administrators, but not hosting
providers or website owners. If a user's privacy goal is to conceal their
online activities from hosting providers and web owners, browsing via
privacy-enhancing network relays (e.g., Tor~\cite{Tor04}) would be a more
suitable option.

\section{Related Work}
\label{sec:related}

Encrypted network traffic analysis has gained more attention in recent years
as the Internet is on its way to be fully encrypted since obtaining a TLS
certificate has become free of charge and easier than
ever~\cite{EFF_ecrypt_theWeb}. However, some initial concept of this class of
attack has been established since the 90s~\cite{Wagner:WOEC96}. As one of the
first attempts to apply traffic analysis on WF, Andrew Hintz simply counts the
number of downloaded files and their size based on the number of connections,
generating fingerprints for a set of targeted websites~\cite{HintzPET2002}.
Similarly, Sun et al.~\cite{Sun:SP02} conduct a large-scale study on the
fingerprintability of 100K webpages based on the number of objects requested
as part of each website's download. However, this attack vector is no longer
effective due to the introduction of persistent HTTP (default since
HTTP/1.1~\cite{rfc2616}) in which multiple files can be transmitted over a
single TCP connection.

In the same year of the two studies above, the pre-alpha version of Tor was
released~\cite{Tor_alpha}, bringing online privacy to another level by not
only encrypting the network traffic but also hiding the fact that a Tor user
is browsing a particular website from both local network observers and the
remote web server~\cite{Tor04}. Since then, the literature has witnessed
numerous studies on WF attacks on Tor using various techniques, ranging from
classical machine learning methods~\cite{Liberatore2006CCS, Wang2014USEC,
Panchenko2016NDSS, Hayes:USEC2016} to advanced deep neural
networks~\cite{Nasr:CCS18, Sirinam:CCS2018}.

Similar to any other privacy-enhancing communications (e.g., Tor), encrypted
DNS traffic is susceptible to traffic analysis. Therefore, padding was added
to remedy this problem~\cite{rfc8467}. However, recent studies find that
current padding strategies are not sufficient to cope with traffic analysis.
Bushart et al.~\cite{bushart_padding} show that padded encrypted DNS traffic
is still vulnerable to traffic analysis attacks. Based on the size and timing
information of encrypted DNS packets, the authors could deanonymize 86.1\% of
10K websites studied. Using the sequence of bytes as a key feature to build a
model for classifying encrypted DoH traffic, Siby et al.~\cite{Siby:NDSS20}
could obtain a precision of 94\% on a dataset of 5K domains. In another
related work, Houser et al.~\cite{Houser:2019:CoNEXT19} analyze DoT traffic
using a classifier based on numerous statistical features extracted from the
time of DNS packets, obtaining an accuracy of 83\% for a dataset of 98
websites. Compared to the scale of our measurement, these prior studies employ
several machine learning techniques on much smaller datasets, with the largest
open-world dataset comprising only 10K domain names~\cite{bushart_padding}.

When WF attacks are designed based primarily on traffic features, such as
packet size and burst, they can be thwarted by obfuscating or adding noise to
the traffic, as is evident by a series of defensive techniques for Tor
proposed previously~\cite{Luo:NDSS2011, Nithyanand:WPES14, Cai2014CCS,
Juarez:ESORICS2016, Cherubin:PETS17, Wang2017USEC, Gong:USEC20}. Siby et
al.~\cite{Siby:NDSS20} has indeed come to a conclusion that routing DoH
traffic via Tor can effectively mitigate their WF attack. There have been
several implementations of DoH over Tor~\cite{cloudflareDNS-Tor, Muffett2021},
which can help to remedy the situation. This is the fundamental reason why we
care about fingerprinting at the IP level, and refrain from using other
traffic features. Specifically, while DoT/DoH traffic can be obfuscated by
tunneling via Tor to cope with these prior attacks, our attack does not target
the DoT/DoH traffic itself but the actual destination IPs contacted when a
website is visited, which are more challenging to hide or obfuscate. One may
suggest the use of Tor in this case as a countermeasure. Nonetheless, it is
important to stress that the fundamental privacy risk that domain encryption
techniques aim to address is orthogonal to those of Tor.

In terms of attacks using the IP address information, Hoang et
al.~\cite{Hoang2020:ASIACCS} assess the privacy benefits offered by domain
name encryption by simply resolving domains into IPs and estimate their
co-location degree without actually visiting any websites. The authors
conclude that co-hosting can help to improve privacy. While this observation
is valid, our WF method could still achieve a high accuracy rate regardless of
many co-hosted websites (see Appendix~\ref{sec:secondary_analyses} for
details). Martino et al.~\cite{Martino2020KnockingOI} conducted a similar
study and could convert IP addresses to their associated domains for the
Tranco top 6K websites with an accuracy of 50.5\%. Patil et
al.~\cite{Patil:ANRW2019} conduct a one-off measurement study to examine the
uniqueness of IP-based fingerprints and find that 95.7\% of websites have a
unique fingerprint. However, similar to most prior WF studies, they do not
consider the impact of caching while also lacking the temporal aspect of
fingerprints. In practice, these essential factors cannot be neglected because
the dynamics of web contents~\cite{levene2004web, Baeza-Yates2004} and
domain--IP mappings over time~\cite{Hoang2020:ASIACCS} can impact the
fingerprints~\cite{Mariano:ARES19}. We address these shortcomings by not only
taking browser caching into consideration but also conducting our measurement
in a longitudinal fashion to investigate the effectiveness of our fingerprints
over time.
\vspace{-1.5em}

\section{Conclusion}
\label{sec:conclusion}
\vspace{-.5em}

Domain name encryption is an important and necessary step to safeguard the
domain name information, which is still largely being transferred in an
unsecured manner on the Internet. However, we have shown that encryption alone
is not enough to protect web users' privacy. Especially when it comes to
preventing nosy network observers from tracking users' browsing
activities, the IP address information of remote servers being contacted is
still visible, which can then be employed to infer the visited websites.

In this study, we construct IP-based fingerprints for more than 200K websites
by performing active DNS measurement to periodically resolve the contacted domain
names while visiting these websites. Using these IP-based
fingerprints, we could successfully identify 84\% of the websites based
solely on the IP addresses observed from the network traffic. Even when browser
caching or ad blocking is considered, reducing the network traffic an
on-path adversary can observe, our fingerprinting technique can still identify
80\% of the websites studied.

Our findings show that significant effort still needs to be invested by both
website owners and hosting providers to maximize the privacy benefits offered
by domain name encryption. Specifically, website owners should try to minimize
references to web resources loaded from domains other than their website's
primary domain, and refrain from hosting their website on servers that do not
co-host any other websites. Hosting providers can help to hinder IP-based WF
by co-locating many websites on the same server(s), while also dynamically
changing mappings between domains and their hosting IPs.
\vspace{-1.5em}

\section*{Acknowledgments}
We would like to thank our shepherd, Tobias Pulls, and the anonymous
reviewers for their thorough feedback on earlier drafts of this paper. We also
thank Shachee Mishra, Tapti Palit, Seyedhamed Ghavamnia, Jarin Firose Moon, Md
Mehedi Hasan, Kien Huynh, Thang Bui, Huan Nguyen, and Thai Le for helpful
conversations and suggestions.

This research was supported in part by the Open Technology Fund under an
Information Controls Fellowship. The opinions in this paper are those of the
authors and do not necessarily reflect the opinions of the sponsor.

{\footnotesize
\bibliographystyle{plain}
\bibliography{main}

\begin{thebibliography}{100}

\bibitem{Tor_alpha}
{Pre-alpha: Run an Onion Proxy Now!}
\newblock
  \url{https://lists.torproject.org/pipermail/tor-dev/2002-September/002374.html}.

\bibitem{NetFlow}
{Cisco IOS NetFlow}.
\newblock \url{http://bit.ly/CiscoNetFlow}, 2012.

\bibitem{EFF_ecrypt_theWeb}
{Encrypt the Web}.
\newblock https://eff.org/encrypt-the-web, 2019.

\bibitem{cloudflare_DoH}
{Cloudflare DoH}.
\newblock http://bit.ly/CloudflareDoH, 2020.

\bibitem{quantcast}
Quantcast.
\newblock https://www.quantcast.com/top-sites/, 2020.

\bibitem{statcounter}
{Stat Counter: Browser Market Share Worldwide}.
\newblock https://gs.statcounter.com/browser-market-share, 2020.

\bibitem{httpArchive}
{State of the Web}.
\newblock https://httparchive.org/reports/state-of-the-web, 2020.

\bibitem{verisign.domains}
Verisign report - the domain name industry brief.
\newblock https://bit.ly/Verisign-Report, 2020.

\bibitem{alexa}
{Alexa Top Sites}.
\newblock https://www.alexa.com/, 2021.

\bibitem{firehol}
{IP Feeds by FireHOL}.
\newblock \url{https://iplists.firehol.org/}, 2021.

\bibitem{majestic}
{The Majestic Million}.
\newblock http://bit.ly/MajesticList, 2021.

\bibitem{cisco_umbrella}
Umbrella popularity list.
\newblock http://bit.ly/UmbrellaList, 2021.

\bibitem{Akbari2019PlatformSA}
Azadeh Akbari and Rashid Gabdulhakov.
\newblock {Platform Surveillance and Resistance in Iran and Russia : The Case
  of Telegram}.
\newblock In {\em {Surveillance and Society}}, 2019.

\bibitem{Tripletcensors}
Anonymous, AA. Niaki, NP. Hoang, P.~Gill, and A.~Houmansadr.
\newblock Triplet censors: Demystifying great firewall{\textquoteright}s {DNS}
  censorship behavior.
\newblock In {\em {USENIX FOCI '20}}.

\bibitem{Baeza-Yates2004}
Ricardo Baeza-Yates, Carlos Castillo, and Felipe Saint-Jean.
\newblock {\em {Web Dynamics, Structure, and Page Quality}}.
\newblock 2004.

\bibitem{Beverly2004:PAM}
Robert Beverly.
\newblock {A Robust Classifier for Passive TCP/IP Fingerprinting}.
\newblock In {\em PAM '04}.

\bibitem{rfc3546}
Simon Blake-Wilson, Magnus Nystrom, David Hopwood, Jan Mikkelsen, and Tim
  Wright.
\newblock {Transport Layer Security (TLS) Extensions}.
\newblock RFC 3546, IETF, June 2003.

\bibitem{ISP_Forbes}
Thomas Brewster.
\newblock {Now Those Privacy Rules Are Gone, This Is How ISPs Will Actually
  Sell Your Personal Data}.
\newblock \url{https://bit.ly/Forbes-ISP-sells-data}, 2017.

\bibitem{bushart_padding}
J.~Bushart and C.~Rossow.
\newblock Padding ain{\textquoteright}t enough: Assessing the privacy
  guarantees of encrypted {DNS}.
\newblock In {\em {FOCI '20}}.

\bibitem{Cai2014CCS}
X.~Cai, R.~Nithyanand, T.~Wang, R.~Johnson, and I.~Goldberg.
\newblock {A Systematic Approach to Developing and Evaluating Website
  Fingerprinting Defenses}.
\newblock In {\em {ACM CCS '14}}.

\bibitem{Cangialosi2016CCS}
Frank Cangialosi, Taejoong Chung, David Choffnes, Dave Levin, Bruce~M. Maggs,
  Alan Mislove, and Christo Wilson.
\newblock Measurement and analysis of private key sharing in the https
  ecosystem.
\newblock In {\em {ACM CCS '16}}.

\bibitem{Cherubin:PETS17}
Giovanni Cherubin, Jamie Hayes, and Marc Juarez.
\newblock {Website Fingerprinting Defenses at the Application Layer}.
\newblock 2017.

\bibitem{Coull2007OnWB}
S.~Coull, M.~Collins, C.~Wright, F.~Monrose, and M.~Reiter.
\newblock On web browsing privacy in anonymized netflows.
\newblock In {\em USENIX Security '07}.

\bibitem{Cui2019RevisitingAF}
W.~Cui, T.~Chen, C.~Fields, J.~Chen, A.~Sierra, and E.~Chan-Tin.
\newblock Revisiting assumptions for website fingerprinting attacks.
\newblock In {\em {ACM AsiaCCS '19}}.

\bibitem{Deccio:CoNEXT19}
Casey Deccio and Jacob Davis.
\newblock {DNS Privacy in Practice and Preparation}.
\newblock In {\em {ACM CoNEXT '19}}.

\bibitem{Tor04}
R.~Dingledine, N.~Mathewson, and P.~Syverson.
\newblock {Tor: The Second-Generation Onion Router}.
\newblock In {\em {USENIX Security '04}}.

\bibitem{holdonDNS}
H.~Duan, N.~Weaver, Z.~Zhao, M.~Hu, J.~Liang, J.~Jiang, K.~Li, and V.~Paxson.
\newblock {Hold-On: Protecting Against On-Path DNS Poisoning}.
\newblock In {\em {SATIN '12}}.

\bibitem{Felt2017MeasuringHA}
AP. Felt, R.~Barnes, A.~King, C.~Palmer, C.~Bentzel, and P.~Tabriz.
\newblock {Measuring HTTPS Adoption on the Web}.
\newblock In {\em USENIX Security '17}.

\bibitem{rfc2616}
R.~Fielding, J.~Gettys, J.~Mogul, H.~Frystyk, L.~Masinter, P.~Leach, and
  T.~Berners-Lee.
\newblock {HTTP/1.1}.
\newblock RFC 2616, June 1999.

\bibitem{Forgy1965ClusterAO}
Edward~W. Forgy.
\newblock {Cluster Analysis of Multivariate Data : Efficiency Versus
  Interpretability of Classifications}.
\newblock 1965.

\bibitem{Fuchs2011InternetAS}
Christian Fuchs, Kees Boersma, Anders Albrechtslund, and Marisol Sandoval.
\newblock {Internet and Surveillance: The Challenges of Web 2.0 and Social
  Media}.
\newblock 2011.

\bibitem{Goldberg:COMPCON97}
I.~Goldberg, D.~Wagner, and E.~Brewer.
\newblock {Privacy-Enhancing Technologies for the Internet}.
\newblock In {\em Proceedings of the 42nd IEEE International Computer
  Conference}, 1997.

\bibitem{Gong:USEC20}
J.~Gong and T.~Wang.
\newblock {Zero-delay Lightweight Defenses against Website Fingerprinting}.
\newblock In {\em {USENIX Security '20'}}.

\bibitem{Gonzalez2016UserPI}
R.~Gonzalez, Claudio Soriente, and Nikolaos Laoutaris.
\newblock User profiling in the time of https.
\newblock In {\em {ACM IMC '16}}.

\bibitem{googleDoH}
Google.
\newblock {JSON API for DNS over HTTPS (DoH)}.
\newblock https://developers.google.com/speed/public-dns/docs/dns-over-https,
  2019.

\bibitem{BlockingJS}
{Google Developers}.
\newblock {Remove Render-Blocking JavaScript}.
\newblock \url{https://developers.google.com/speed/docs/insights/BlockingJS},
  2018.

\bibitem{Greschbach2017TheEO}
B.~Greschbach, T.~Pulls, LM. Roberts, P.~Winter, and N.~Feamster.
\newblock {The Effect of DNS on Tor's Anonymity}.
\newblock In {\em {NDSS '17}}.

\bibitem{Grigorik:2018}
Ilya Grigorik.
\newblock {Critical Rendering Path}.
\newblock http://bit.ly/CriticalRenderingPath, 2018.

\bibitem{ChinaBansVPN}
B.~Haas.
\newblock {Man in China Sentenced to Five years' Jail for Running VPN}.
\newblock
  \url{https://www.theguardian.com/world/2017/dec/22/man-in-china-sentenced-to-five-years-jail-for-running-vpn}.

\bibitem{Hayes:USEC2016}
J.~Hayes and G.~Danezis.
\newblock {k-fingerprinting: A Robust Scalable Website Fingerprinting
  Technique}.
\newblock In {\em {USENIX Security Symposium 2016}}.

\bibitem{HintzPET2002}
A.~Hintz.
\newblock {Fingerprinting Websites Using Traffic Analysis}.
\newblock In {\em Conference on Privacy Enhancing Technologies}, 2002.

\bibitem{Hoang2017:locationPrivacy}
NP. Hoang, Y.~Asano, and M.~Yoshikawa.
\newblock {Your Neighbors Are My Spies: Location and other Privacy Concerns in
  GLBT-focused Location-based Dating Applications}.
\newblock In {\em {Trans. on Advanced Communications Technology 2016}}.

\bibitem{Hoang2018:IMC}
NP. Hoang, P.~Kintis, M.~Antonakakis, and M.~Polychronakis.
\newblock {An Empirical Study of the I2P Anonymity Network and Its Censorship
  Resistance}.
\newblock In {\em {ACM IMC '18}}.

\bibitem{Hoang2020:MADWeb}
NP. Hoang, I.~Lin, S.~Ghavamnia, and M.~Polychronakis.
\newblock {K-resolver: Towards Decentralizing Encrypted DNS Resolution}.
\newblock In {\em MADWeb '20}.

\bibitem{Hoang2020:ASIACCS}
NP. Hoang, AA. Niaki, N.~Borisov, P.~Gill, and M.~Polychronakis.
\newblock {Assessing the Privacy Benefits of Domain Name Encryption}.
\newblock In {\em {ACM AsiaCCS '20}}.

\bibitem{GFWatch}
NP. Hoang, AA. Niaki, J.~Dalek, J.~Knockel, P.~Lin, B.~Marczak,
  M.~Crete-Nishihata, P.~Gill, and M.~Polychronakis.
\newblock {How Great is the Great Firewall? Measuring China's DNS Censorship}.
\newblock In {\em {USENIX Security '21}}.

\bibitem{Hoang2020:WebCo-location}
NP. Hoang, AA. Niaki, M.~Polychronakis, and P.~Gill.
\newblock {The Web is Still Small After More Than a Decade}.
\newblock {\em {ACM SIGCOMM Computer Communication Review 2020}}.

\bibitem{Hoang2014:Anonymous}
NP. Hoang and D.~Pishva.
\newblock {Anonymous Communication and Its Importance in Social Networking}.
\newblock In {\em {ICACT '14}}.

\bibitem{rfc8484}
P.~Hoffman and P.~McManus.
\newblock {DNS} queries over {HTTPS}.
\newblock RFC 8484, IETF, October 2018.

\bibitem{Houser:2019:CoNEXT19}
Rebekah Houser, Zhou Li, Chase Cotton, and Haining Wang.
\newblock {An Investigation on Information Leakage of DNS over TLS}.
\newblock In {\em {ACM CoNEXT}}, 2019.

\bibitem{rfc7858}
Z.~Hu, L.~Zhu, J.~Heidemann, A.~Mankin, D.~Wessels, and P.~Hoffman.
\newblock Specification for {DNS} over transport layer security ({TLS}).
\newblock RFC 7858, IETF, May 2016.

\bibitem{firefoxECH}
Kevin Jacobs.
\newblock {Encrypted Client Hello: the future of ESNI in Firefox}.
\newblock
  \url{http://blog.mozilla.org/security/2021/01/07/encrypted-client-hello-the-future-of-esni-in-firefox},
  2021.

\bibitem{Juarez:CCS2014}
Marc Juarez, Sadia Afroz, Gunes Acar, Claudia Diaz, and Rachel Greenstadt.
\newblock {A Critical Evaluation of Website Fingerprinting Attacks}.
\newblock In {\em {ACM CCS}}, 2014.

\bibitem{Juarez:ESORICS2016}
Marc Juarez, Mohsen Imani, Mike Perry, Claudia Diaz, and Matthew Wright.
\newblock Toward an efficient website fingerprinting defense.
\newblock In {\em ESORICS}, 2016.

\bibitem{ISP_wsj}
Sarah Krouse and Patience Haggin.
\newblock {Internet Providers Look to Cash In on Your Web Habits}.
\newblock
  \url{https://www.wsj.com/articles/facebook-knows-a-lot-about-you-so-does-your-internet-provider-11561627803},
  2019.

\bibitem{Leith2020WebBP}
Douglas~J. Leith.
\newblock Web browser privacy: What do browsers say when they phone home?
\newblock 2020.

\bibitem{levene2004web}
Mark Levene.
\newblock {\em Web dynamics: Adapting to change in content, size, topology and
  use}.
\newblock Springer Science \& Business Media, 2004.

\bibitem{Liberatore2006CCS}
M.~Liberatore and BN. Levine.
\newblock {Inferring the Source of Encrypted HTTP Connections}.
\newblock In {\em {ACM CCS '06}}, 2006.

\bibitem{Libert2019GoodNF}
T.~Libert and R.~Binns.
\newblock Good news for people who love bad news: Centralization, privacy, and
  transparency on us news sites.
\newblock {\em ACM Conference on Web Science}, 2019.

\bibitem{Lloyd1982LeastSQ}
Stuart~P. Lloyd.
\newblock Least squares quantization in pcm.
\newblock 1982.

\bibitem{Lu:2019:IMC19}
Chaoyi Lu, Baojun Liu, Zhou Li, Shuang Hao, Haixin Duan, Mingming Zhang,
  Chunying Leng, Ying Liu, Zaifeng Zhang, and Jianping Wu.
\newblock {An End-to-End, Large-Scale Measurement of DNS-over-Encryption: How
  Far Have We Come?}
\newblock In {\em {ACM Internet Measurement Conference}}, 2019.

\bibitem{Luo:NDSS2011}
X.~Luo, P.~Zhou, E.~Chan, W.~Lee, R.~Chang, and R.~Perdisci.
\newblock {HTTPOS: Sealing Information Leaks with Browser-side Obfuscation of
  Encrypted Flows}.
\newblock In {\em {Network and Distributed System Security Symposium}}, 2011.

\bibitem{MacQueen1967SomeMF}
James~B. MacQueen.
\newblock Some methods for classification and analysis of multivariate
  observations.
\newblock 1967.

\bibitem{Mariano:ARES19}
M.~Di Martino, P.~Quax, and W.~Lamotte.
\newblock {Realistically Fingerprinting Social Media Webpages in HTTPS
  Traffic}.
\newblock In {\em {ACM ARES '19}}.

\bibitem{Martino2020KnockingOI}
Mariano~Di Martino, P.~Quax, and W.~Lamotte.
\newblock {Knocking on IPs: Identifying HTTPS Websites for Zero-Rated Traffic}.
\newblock {\em {Security and Communication Networks 2020}}.

\bibitem{rfc8467}
A.~Mayrhofer.
\newblock {Padding Policies for EDNS(0)}.
\newblock RFC 8467, IETF, 2018.

\bibitem{firefoxDoH_start}
Patrick McManus.
\newblock Improving {DNS} privacy in firefox.
\newblock
  \url{https://blog.nightly.mozilla.org/2018/06/01/improving-dns-privacy-in-firefox/},
  2018.

\bibitem{Domain_sharding}
{MDN Web Docs}.
\newblock {Domain sharding}.
\newblock
  \url{https://developer.mozilla.org/en-US/docs/Glossary/Domain\_sharding},
  2020.

\bibitem{DOMContentLoaded}
{MDN Web Docs}.
\newblock {DOMContentLoaded event}.
\newblock
  \url{https://developer.mozilla.org/en-US/docs/Web/API/Window/DOMContentLoaded\_event},
  2020.

\bibitem{Miller:PETS14}
Brad Miller, Ling Huang, Anthony~D. Joseph, and J.~Doug Tygar.
\newblock {I Know Why You Went to the Clinic: Risks and Realization of HTTPS
  Traffic Analysis}.
\newblock In {\em {Privacy Enhancing Technologies Symposium}}, 2014.

\bibitem{Muffett2021}
Alec Muffett.
\newblock {No Port 53, Who Dis? A year of DNS over HTTPS over Tor}.
\newblock In {\em {DNS Privacy Workshop 2021}}.

\bibitem{IndiaJail}
Rayan Naqash.
\newblock {India's crackdown on VPNs in Kashmir seeks to quell cyber-insurgency
  threat but risks blowback}.
\newblock \url{hhttps://bit.ly/India-blocks-VPN}, 2020.

\bibitem{Nasr:CCS18}
Milad Nasr, Alireza Bahramali, and Amir Houmansadr.
\newblock {DeepCorr: Strong Flow Correlation Attacks on Tor Using Deep
  Learning}.
\newblock In {\em {ACM CCS}}, 2018.

\bibitem{Nasr2017CompressiveTA}
Milad Nasr, Amir Houmansadr, and A.~Mazumdar.
\newblock Compressive traffic analysis: A new paradigm for scalable traffic
  analysis.
\newblock In {\em {ACM CCS '17}}.

\bibitem{iclab_SP20}
Arian~Akhavan Niaki, Shinyoung Cho, Zachary Weinberg, Nguyen~Phong Hoang, Abbas
  Razaghpanah, Nicolas Christin, and Phillipa Gill.
\newblock {{ICLab}: A Global, Longitudinal Internet Censorship Measurement
  Platform}.
\newblock In {\em {Symposium on Security and Privacy}}, May 2020.

\bibitem{Nikiforakis2012}
Nick Nikiforakis, Luca Invernizzi, Alexandros Kapravelos, Steven Van~Acker,
  Wouter Joosen, Christopher Kruegel, Frank Piessens, and Giovanni Vigna.
\newblock You are what you include: Large-scale evaluation of remote javascript
  inclusions.
\newblock In {\em {ACM Conference on Computer and Communications Security}},
  2012.

\bibitem{Nithyanand:WPES14}
Rishab Nithyanand, Xiang Cai, and Rob Johnson.
\newblock Glove: A bespoke website fingerprinting defense.
\newblock In {\em {WPES}}, 2014.

\bibitem{hoang:2019:measuringI2P}
{NP. Hoang and S. Doreen and M. Polychronakis}.
\newblock {Measuring I2P Censorship at a Global Scale}.
\newblock In {\em {FOCI '19}}.

\bibitem{Panchenko2016NDSS}
Andriy Panchenko, Fabian Lanze, Jan Pennekamp, Thomas Engel, Andreas Zinnen,
  Martin Henze, and Klaus Wehrle.
\newblock Website fingerprinting at internet scale.
\newblock In {\em {Network and Distributed System Security Symposium}}, 2016.

\bibitem{Panchenko2011WPES}
Andriy Panchenko, Lukas Niessen, Andreas Zinnen, and Thomas Engel.
\newblock Website fingerprinting in onion routing based anonymization networks.
\newblock In {\em {WPES}}, 2011.

\bibitem{Patil:ANRW2019}
S.~Patil and N.~Borisov.
\newblock {What Can You Learn from an IP?}
\newblock In {\em {Applied Networking Research Workshop}}, 2019.

\bibitem{Pearce:2017:Iris}
Paul Pearce, Ben Jones, Frank Li, Roya Ensafi, Nick Feamster, Nick Weaver, and
  Vern Paxson.
\newblock {Global Measurement of {DNS} Manipulation}.
\newblock In {\em {USENIX Security '17}}, 2017.

\bibitem{TorCritique}
Mike Perry.
\newblock {A Critique of Website Traffic Fingerprinting Attacks}, 2013.
\newblock
  \url{https://blog.torproject.org/critique-website-traffic-fingerprinting-attacks}.

\bibitem{LePochat:CSET2019}
Victor~Le Pochat, Tom~Van Goethem, and Wouter Joosen.
\newblock Evaluating the long-term effects of parameters on the characteristics
  of the tranco top sites ranking.
\newblock In {\em {USENIX Workshop on Cyber Security Experimentation and
  Test}}, 2019.

\bibitem{Pulls:PETS20}
Tobias Pulls and Rasmus Dahlberg.
\newblock Website fingerprinting with website oracles.
\newblock {\em {PETS}}, 2020.

\bibitem{Razaghpanah2018AppsTP}
Abbas Razaghpanah, Rishab Nithyanand, Narseo Vallina-Rodriguez, Srikanth
  Sundaresan, Mark Allman, Christian Kreibich, and Phillipa Gill.
\newblock Apps, trackers, privacy, and regulators: A global study of the mobile
  tracking ecosystem.
\newblock In {\em {Network and Distributed System Security Symposium}}, 2018.

\bibitem{rfc-draft-ietf-tls-esni-06}
E.~Rescorla, K.~Oku, N.~Sullivan, and C.~Wood.
\newblock {ESNI for {TLS} 1.3}.
\newblock Internet draft, IETF, March 2020.

\bibitem{rfc-draft-ietf-tls-esni-07}
E.~Rescorla, K.~Oku, N.~Sullivan, and C.~Wood.
\newblock {TLS Encrypted Client Hello}.
\newblock Internet draft, IETF, June 2020.

\bibitem{Rweyemamu2019}
Walter Rweyemamu, Christo Lauinger, Tobiasand~Wilson, William Robertson, and
  Engin Kirda.
\newblock {Clustering and the Weekend Effect: Recommendations for the Use of
  Top Domain Lists in Security Research}.
\newblock In {\em {PAM}}, 2019.

\bibitem{rfc6960}
S.~Santesson, M.~Myers, R.~Ankney, A.~Malpani, S.~Galperin, and C.~Adams.
\newblock {X.509 Internet Public Key Infrastructure Online Certificate Status
  Protocol - OCSP}.
\newblock RFC 6960, IETF, June 2013.

\bibitem{cloudflareDNS-Tor}
Mahrud Sayrafi.
\newblock Introducing {DNS} resolver for {T}or.
\newblock \url{https://blog.cloudflare.com/welcome-hidden-resolver/}, 2018.

\bibitem{schmitt_oblivious}
Paul Schmitt, Anne Edmundson, Allison Mankin, and Nick Feamster.
\newblock {Oblivious DNS: Practical Privacy for DNS Queries}.
\newblock In {\em {PETS}}, 2019.

\bibitem{Shamsi2016HershelSO}
Zain Shamsi, Ankur Nandwani, D.~Leonard, and D.~Loguinov.
\newblock Hershel: Single-packet os fingerprinting.
\newblock {\em IEEE/ACM Transactions on Networking}, 24:2196--2209, 2016.

\bibitem{Shannon:Entropy}
C.~E. Shannon.
\newblock A mathematical theory of communication.
\newblock {\em SIGMOBILE Mob. Comput. Commun. Rev.}, 2001.

\bibitem{Shue:2007}
Craig~A. Shue, Andrew~J. Kalafut, and Minaxi Gupta.
\newblock {The Web is Smaller Than It Seems}.
\newblock In {\em {IMC'07}}.

\bibitem{Siby:NDSS20}
Sandra {Siby}, Marc {Juarez}, Claudia {Diaz}, Narseo {Vallina-Rodriguez}, and
  Carmela {Troncoso}.
\newblock {Encrypted DNS => Privacy? A Traffic Analysis Perspective}.
\newblock In {\em {NDSS '20}}.

\bibitem{Singanamalla2021}
S.~Singanamalla, Suphanat Chunhapanya, Marek Vavrusa, Tanya Verma, P.~Wu,
  Marwan Fayed, K.~Heimerl, N.~Sullivan, and C.~Wood.
\newblock Oblivious dns over https (odoh): A practical privacy enhancement to
  dns.
\newblock In {\em {DNS Privacy Workshop 2021}}.

\bibitem{Sirinam:CCS2018}
Payap Sirinam, Mohsen Imani, Marc Juarez, and Matthew Wright.
\newblock Deep fingerprinting: Undermining website fingerprinting defenses with
  deep learning.
\newblock In {\em {ACM Conference on Computer and Communications Security}},
  2018.

\bibitem{Sun:SP02}
Qixiang Sun, Daniel~R. Simon, Yi-Min Wang, Wilf Russell, Venkata~N.
  Padmanabhan, and Lili Qiu.
\newblock Statistical identification of encrypted web browsing traffic.
\newblock In {\em {IEEE Symposium on Security and Privacy}}, 2002.

\bibitem{Szurdi:usenixsecurity14}
Janos Szurdi, Balazs Kocso, Gabor Cseh, Jonathan Spring, Mark Felegyhazi, and
  Chris Kanich.
\newblock {The Long {\textquotedblleft}Taile{\textquotedblright} of
  Typosquatting Domain Names}.
\newblock In {\em {USENIX Security '14}}.

\bibitem{rfc7015}
B.~Trammell, A.~Wagner, and B.~Claise.
\newblock {Flow Aggregation for the IP Flow Information Export (IPFIX)
  Protocol}.
\newblock RFC 7015, IETF, Sep 2013.

\bibitem{Trevisan2016TowardsWS}
Martino Trevisan, Idilio Drago, Marco Mellia, and Maurizio~M. Munaf{\`o}.
\newblock Towards web service classification using addresses and dns.
\newblock In {\em {IWCMC}}, 2016.

\bibitem{Trevisan2020DoesDN}
Martino Trevisan, Francesca Soro, M.~Mellia, I.~Drago, and Ricardo Morla.
\newblock Does domain name encryption increase users' privacy?
\newblock {\em ACM SIGCOMM Computer Communication Review}, 50:16 -- 22, 2020.

\bibitem{LePochat2019}
{V. Le Pochat and T. Van Goethem and S. Tajalizadehkhoob and M. Korczy\'{n}ski
  and W. Joosen}.
\newblock {Tranco: A Research-Oriented Top Sites Ranking Hardened Against
  Manipulation}.
\newblock In {\em NDSS '19}.

\bibitem{Verde2014NoNU}
Nino~Vincenzo Verde, G.~Ateniese, E.~Gabrielli, L.~Mancini, and A.~Spognardi.
\newblock No nat'd user left behind: Fingerprinting users behind nat from
  netflow records alone.
\newblock {\em 2014 IEEE 34th International Conference on Distributed Computing
  Systems}, pages 218--227, 2014.

\bibitem{Wagner:WOEC96}
David Wagner and Bruce Schneier.
\newblock Analysis of the ssl 3.0 protocol.
\newblock In {\em {Workshop on Electronic Commerce}}, 1996.

\bibitem{Wang2020SP}
Tao Wang.
\newblock High precision open-world website fingerprinting.
\newblock In {\em {IEEE S\&P}}, 2020.

\bibitem{Wang2014USEC}
Tao Wang, Xiang Cai, Rishab Nithyanand, Rob Johnson, and Ian Goldberg.
\newblock Effective attacks and provable defenses for website fingerprinting.
\newblock In {\em {USENIX} Security}, 2014.

\bibitem{Wang2017USEC}
Tao Wang and Ian Goldberg.
\newblock Walkie-talkie: An efficient defense against passive website
  fingerprinting attacks.
\newblock In {\em {USENIX Security}}, 2017.

\bibitem{Wang:NSDI14}
Xiao~Sophia Wang, Aruna Balasubramanian, Arvind Krishnamurthy, and David
  Wetherall.
\newblock How speedy is {SPDY}?
\newblock In {\em {USENIX NSDI}}, 2014.

\bibitem{Xu2018ACSAC}
Yixiao Xu, Tao Wang, Qi~Li, Qingyuan Gong, Yang Chen, and Yong Jiang.
\newblock {A Multi-Tab Website Fingerprinting Attack}.
\newblock In {\em {ACSAC}}, 2018.

\bibitem{ChinaBansGame}
Sophia Yang.
\newblock {China to ban online gaming, chatting with foreigners outside Great
  Firewall: Report}.
\newblock \url{https://www.taiwannews.com.tw/en/news/3916690}, 2020.

\bibitem{petcon2009-zzz}
zzz and Lars Schimmer.
\newblock {Peer Profiling and Selection in the I2P Anonymous Network}.
\newblock In {\em {PET-CON}}, 2009.

\end{thebibliography}
}
\vspace{-1.5em}
\appendix
\section{Dynamics of Domain-IP Mapping}
\label{sec:Domain_ip_dynamics}

In this appendix, we analyze the mappings between domains and IP addresses
observed throughout the whole study period to examine the dynamics of
domain--IP mappings in today's web ecosystem. Over the two-month period, we
observed 531K domain names, resulting in 693K unique IP address. Of these
domains, 212K belong to the primary domain group selected
in~\sectionref{sec:experiment_setup}, and 319K are secondary domains. In
total, we have gathered more than 7M domain--IP mappings.

\begin{figure}
    \centering

    \begin{subfigure}[b]{0.85\columnwidth}
        \includegraphics[width=\columnwidth]{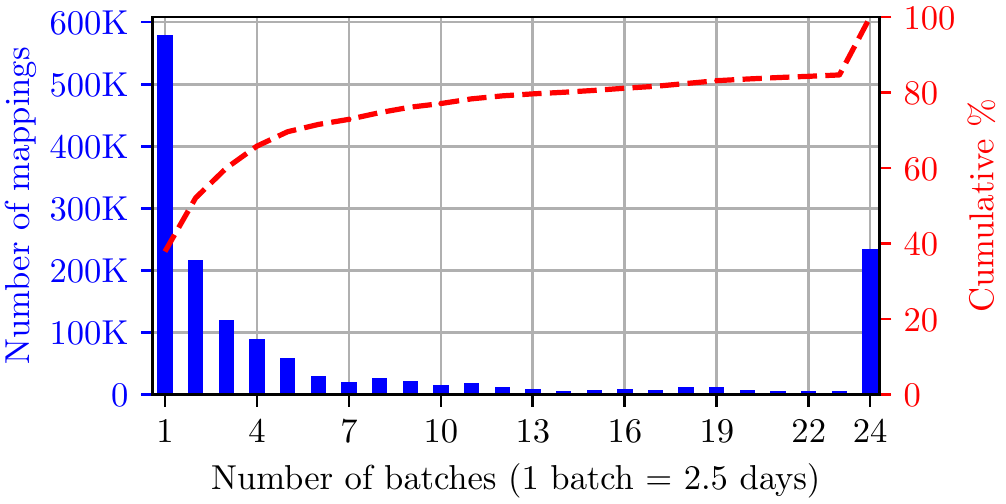}
        \caption{Primary domains}
        \label{fig:longevity_primary_domain}
    \end{subfigure}

    \begin{subfigure}[b]{0.85\columnwidth}
        \includegraphics[width=\columnwidth]{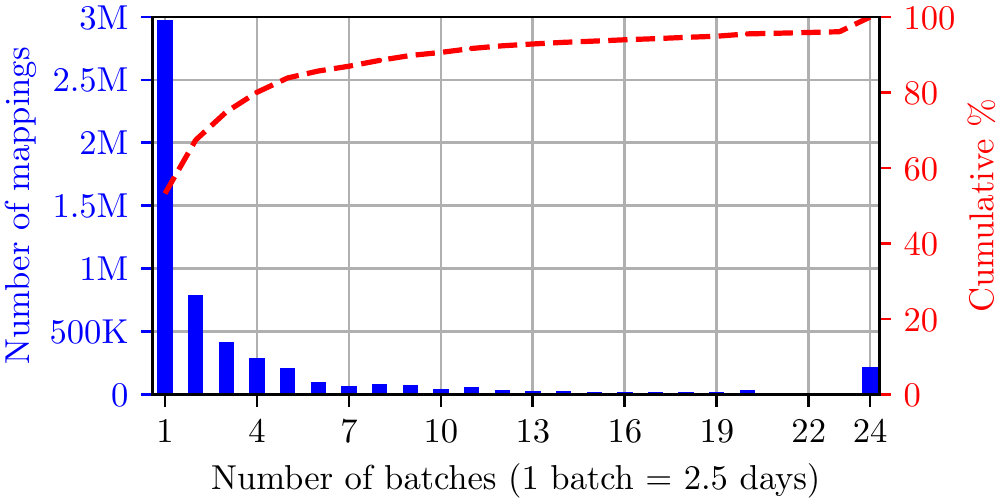}
        \caption{Secondary domains}
        \label{fig:longevity_secondary_domain}
    \end{subfigure}

    \caption{Longevity of mappings between domains and their IPs.}
    \label{fig:mapping_longevity}
\end{figure}

Figure~\ref{fig:mapping_longevity} shows the longevity analysis of domain--IP
mappings of the two domain groups. More than 60\% of the mappings in both
groups last for less than a week (i.e., observed in no more than three
consecutive data batches). In contrast, only 15\% of primary and less than 5\%
of secondary domain mappings can be observed for the whole two-month period of
our study. The high churn rate of most mappings after a week is one of the
reasons why our IP-based fingerprints deteriorate after ten days since being
constructed (\sectionref{sec:fingerprint_stability}).

However, the picture changes completely when examining the number of domains
and IP addresses in each mapping group. We refer to the group of mappings that
are observed in no more than three consecutive data batches as \textit{dynamic
mappings}, and to mappings that are observed continuously for the whole period
of study as \textit{static mappings}.
Table~\ref{tab:mapping_longevity_breakdown} shows the breakdown of the number
of unique domains and IP addresses observed in each mapping group. We can see
two unbalanced allocations between (a) the total number of primary and
secondary domain mappings, and (b) the dynamic and static mappings within the
primary domains.

For (a), the number of secondary domain mappings is almost four times larger
than primary domains, due to the fact that a visit to a primary domain loads
several secondary domains. For (b), there are only 36K (17\%) primary domains
with a high IP address churn rate, occupying a pool of 169K unique IP
addresses. In contrast, 167K (79\%) primary domains remain stable on the same
IP addresses for the whole period of our study. This explains the reason why
many of our IP-based fingerprints are still effective after two months
(\sectionref{sec:fingerprint_stability}). Specifically, although 50\% of the
IP addresses are changed in more than 50\% of the fingerprints, as shown in
Figure~\ref{fig:ip_stability}, this is mainly due to the change of secondary
domains' hosting IP addresses. On the other hand, after two months, almost
80\% of primary domains are still hosted on static IP addresses, contributing
to the validity of our IP-based fingerprints.

\begin{table}[t]
\footnotesize
\centering
\caption{Breakdown of the number of domains and IPs in dynamic and static mappings.}
\begin{tabular}{lrrrr}
\toprule
\textbf{Domain \& IP}& \textbf{Total} &\textbf{Dynamic} & \textbf{Static} \\ [0.5ex]
\textbf{type}     & \textbf{mappings} &\textbf{mappings} & \textbf{mappings} \\ [0.5ex]
\midrule
Primary domains     &            1.5M &      36K (17\%) &             167K (79\%)  \\
Primary IPs         &                 &     169K (45\%) &             137K (36\%)  \\
\midrule
Secondary domains   &            5.5M &     208K (65\%) &              88K (28\%)  \\
Secondary IPs       &                 &     306K (69\%) &              77K (17\%)  \\
\bottomrule
\end{tabular}
\label{tab:mapping_longevity_breakdown}
\end{table}

Although only a small number of domains whose hosting IP addresses are changed
frequently, our findings show that it is totally possible to increase the
dynamics of domain--IP mappings from the perspective of hosting providers. One
may consider that frequently changing the hosting IP addresses is not feasible
for those web servers that use ``cruise-liner''
certificates~\cite{Cangialosi2016CCS}, in which numerous domains are
aggregated in each certificate to support non-SNI clients. However, to the
best of our knowledge, the use of ``cruise-liner'' certificates has been
deprecated by most hosting providers due to widespread support of SNI by all
major browsers. For instance, while Cloudflare used to employ ``cruise-liner''
certificates for websites co-hosted on its CDNs, the Subject Alternative Names
field of Cloudflare's certificates now contains only the websites' domain and
\texttt{sni.cloudflaressl.com}. Thus, our suggestion is still compatible with
multiple certificates per IP address. In fact, Cloudflare does allow website
owners to upload their own certificates instead of using Cloudflare's.

\section{Domains Filtered by Brave}
\label{sec:brave_filtered_domains}

\begin{figure}[t]
    \centering
    \includegraphics[width=0.8\columnwidth]{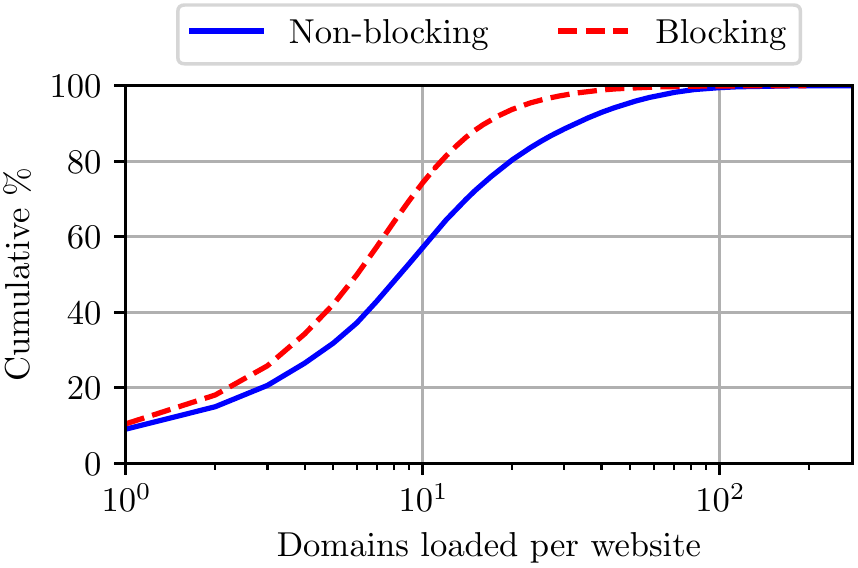}
    \caption{CDF of domain names loaded per website.}
    \label{fig:loaded_domains_per_website}
\end{figure}

\begin{figure}[t]
    \centering
    \includegraphics[width=0.8\columnwidth]{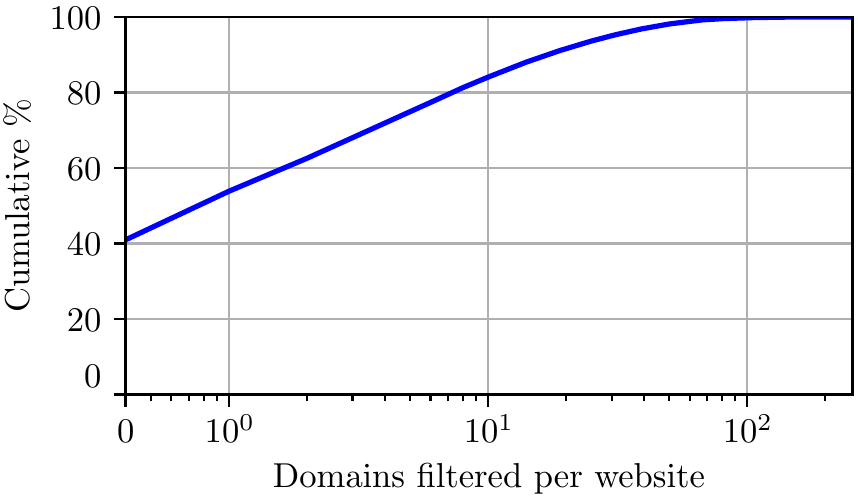}
    \caption{CDF of domains filtered per website by Brave.}
    \label{fig:filtered_domains_per_website}
\end{figure}

While the accuracy rate does decrease when fingerprinting against websites
browsed with Brave (\sectionref{sec:fingerprint_brave}), our basic
fingerprinting approach could still obtain a relatively high accuracy rate of
80\%. To that end, we conduct an additional analysis on the filtered domains
to find the underlying reason why removing these resources does not
substantially reduce websites’ fingerprintability.

As shown in Figure~\ref{fig:loaded_domains_per_website}, the number of domains
loaded per website when browsing with Brave (dashed line) is significantly
lower than when using a non-blocking browser
(solid line). Specifically, almost 80\% of websites load less than
ten domains with Brave, whereas only 57\% of websites load less
than ten domains using a non-blocking browser. The average number
of domains loaded per website with Brave is only
eight, whilst there are 14 domains loaded per website on average for
a non-blocking browser.
Of all websites studied, 41\% of websites do not have any domains filtered by
the Brave browser. The remaining 59\% of these websites have at least one
domain filtered, as shown in Figure~\ref{fig:filtered_domains_per_website}.

Table~\ref{tab:top_domains_removed_brave} shows the top-ten most blocked
domains by Brave, with \texttt{www.google-analytics.com} being the most
blocked domain. Among the 220K websites studied, it is removed from 69K (31\%)
websites. Although the domain is referenced in more than half of the websites
(as shown in Table~\ref{tab:lowest_entropy_domains}), Brave does not entirely
remove it from all of them, depending on how it is referenced on each website.
Despite being removed from a large number of websites, as indicated in the
third column of the table, these domains only contribute a small amount of
information entropy to the fingerprint when included (as discussed in
\sectionref{sec:fingerprint_entropy}). This is the reason why our
fingerprinting technique can still identify a relatively high number (80\%) of
websites when browsing with Brave. Analyzing the MIME type of objects loaded
from these domains, we find that the vast majority of them are used to load
images and scripts used for tracking and advertisement services operated by
Google and Facebook.
\vspace{-1.5em}

\section{Impact of Co-location and Popularity on Attack Accuracy}
\label{sec:secondary_analyses}

\begin{table}[t]
    \footnotesize
    \centering
    \caption{Top-ten domains removed by Brave.}
    \begin{tabular}{lrc}
    \toprule
    \textbf{Domain name}                   & \textbf{\# Blocked}   & \textbf{Entropy} \\  [0.5ex]
    \midrule
    \texttt{www.google-analytics.com}      & 69K (31\%)                  & 0.87             \\
    \texttt{stats.g.doubleclick.net}       & 57K (26\%)                  & 1.53             \\
    \texttt{www.google.com}                & 38K (17\%)                  & 1.44             \\
    \texttt{www.googletagmanager.com}      & 32K (15\%)                  & 1.71             \\
    \texttt{www.facebook.com}              & 31K (14\%)                  & 1.97             \\
    \texttt{googleads.g.doubleclick.net}   & 28K (13\%)                  & 2.10             \\
    \texttt{tpc.googlesyndication.com}     & 21K (10\%)                  & 3.08             \\
    \texttt{connect.facebook.net}          & 21K (10\%)                  & 1.98             \\
    \texttt{adservice.google.com}          & 18K (8\%)                   & 2.68             \\
    \texttt{pagead2.googlesyndication.com} & 18K (8\%)                   & 3.03             \\
    \bottomrule
    \end{tabular}
    \label{tab:top_domains_removed_brave}

\end{table}

\begin{figure*}[h!]
    \centering

    \begin{subfigure}[b]{0.9\textwidth}
        \centering
        \includegraphics[width=\linewidth]{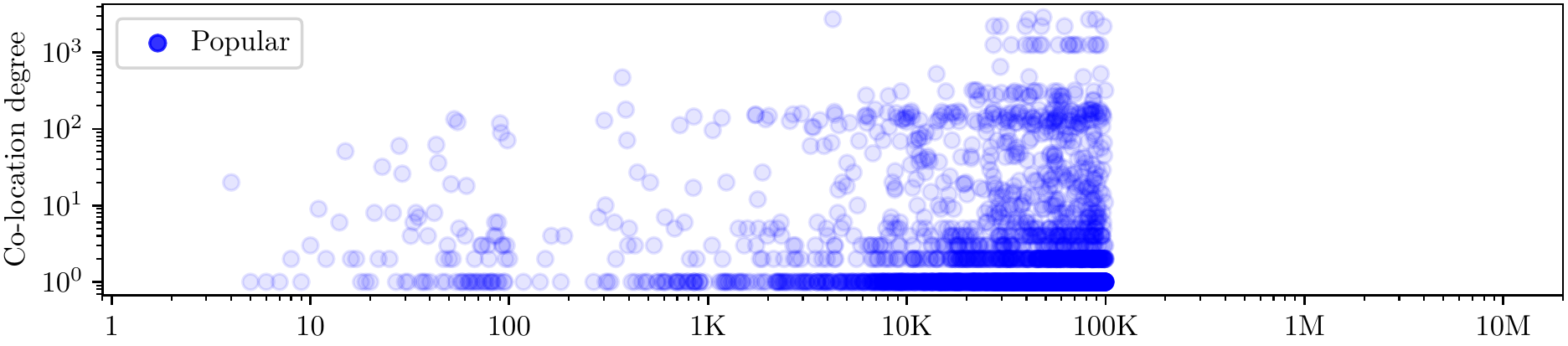}
        \label{fig:fingerprint_accuracy_scatter_popular}
    \end{subfigure}

    \begin{subfigure}[b]{0.9\textwidth}
        \centering
        \includegraphics[width=\linewidth]{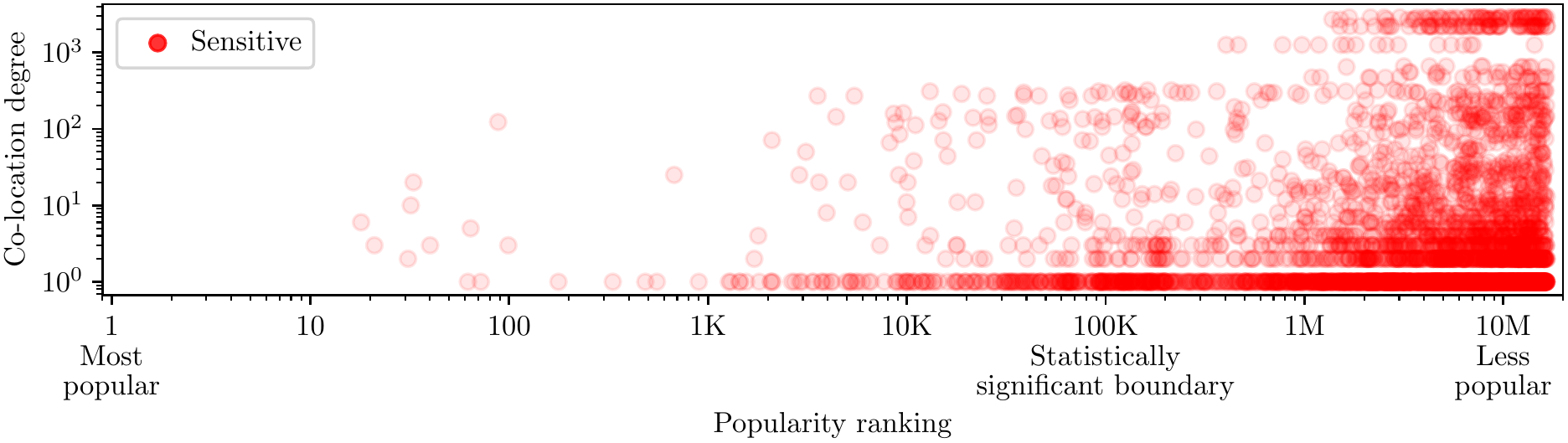}
        \label{fig:fingerprint_accuracy_scatter_sensitive}
    \end{subfigure}
    \caption{Co-location degree vs. popularity ranking distribution of
    \textit{successfully fingerprinted websites}. ``Popular'' corresponds to
    90,231 fingerprinted websites from the Tranco list's top-100K most popular
    domains, while ``Sensitive'' corresponds to 104,983 fingerprinted websites
    from 126,597 sensitive domains chosen from Alexa's sensitive categories
    that span the whole ranking spectrum. 5,687 fingerprinted websites are
    common between the two sets.}
    \label{fig:fingerprint_accuracy_scatter}
\end{figure*}

We next analyze the co-location degree and popularity ranking of the
fingerprinted websites to investigate whether there are any correlations
between these properties of a given website and the chance that it can be
precisely fingerprinted. Figure~\ref{fig:fingerprint_accuracy_scatter} shows
two scatter plots of the popular websites and sensitive websites that we could
successfully fingerprint, with respect to their co-location degree and
popularity ranking. As expected, websites that are not co-hosted with any
other websites are the most susceptible to our IP-based fingerprinting attack.
Regardless of having a high co-location degree, however, websites can still be
fingerprinted with our enhanced technique due to the inclusion of unique
secondary domains.

Hoang et al.~\cite{Hoang2020:ASIACCS} suggest an ideal co-location threshold
of at least 100 domains per hosting IP address, so that the co-hosted websites
can gain some meaningful privacy benefit from the deployment of domain name
encryption. However, Figure~\ref{fig:fingerprint_accuracy_scatter} shows that
even when more than 100 websites are co-hosted, they can still be
fingerprinted. Again, the underlying reason is that these websites often
reference several external resources, making their fingerprint more
distinguishable compared to the rest of the co-hosted websites.

\begin{figure}[!h]
\centering
\includegraphics[width=.9\columnwidth]{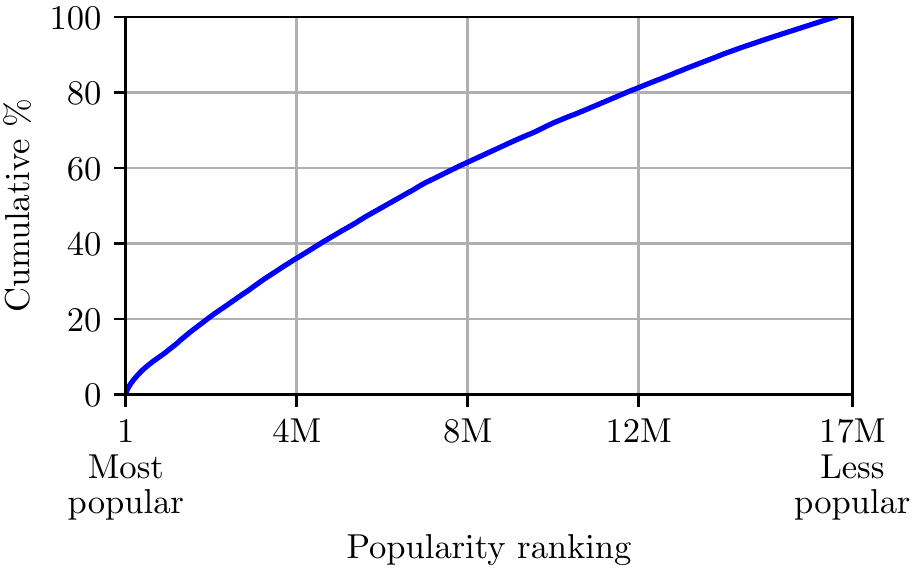}
\caption{CDF of popularity ranking as a percentage of all successfully fingerprinted websites.}
\label{fig:popularity_cdf}
\end{figure}

In addition, Figure~\ref{fig:popularity_cdf} shows the CDF of the popularity
ranking of the successfully fingerprinted websites. We can see that the number
of fingerprinted websites slightly leans towards more popular rankings, which
can also be confirmed by the higher accuracy rates when it comes to
fingerprinting the popular websites compared to the sensitive websites, as
indicated in all three fingerprinting approaches in
Table~\ref{tab:fingerprinting_result_breakdown}.

\end{document}